\documentclass[twoside,twocolumn,english,aps,prb
]{revtex4}
\usepackage{amsmath}
\usepackage{graphicx}
\usepackage{amssymb}

\makeatletter
\usepackage{graphicx}
\usepackage{amsmath}

\usepackage{babel}
\makeatother

\newcommand{\oneoverhbar}{{1 \over \hbar}}
\newcommand{\smalloneoverhbar}{{\textstyle {1 \over \hbar}}}
\newcommand{\oneoverhbarsq}{{1 \over \hbar^2}}

\newcommand{\pp}{\textrm{pp}}
\newcommand{\app}{{\overline { \hspace{-0.1mm} \pp \hspace{-0.3mm}} 
\hspace{0.4mm} }}
\newcommand{\vhbar}{{}}
\newcommand{\toh}{{\textstyle \frac{1}{2}}}

\newcommand{\eg}{\emph{e.g.}} 
\newcommand{\ie}{\emph{i.e.}}
\newcommand{\ve}{\varepsilon}
\newcommand{\tgammat}{\tilde \gamma_t}
\newcommand{\tgammaH}{\tilde \gamma_H}
\newcommand{\vep}{{\varepsilon_+}}
\newcommand{\vem}{{\varepsilon_-}}

\newcommand{\bC}{\bar C}

\newcommand{\tCO}{{\tilde C^0}}
\newcommand{\tP}{{\tilde P}}

\newcommand{\crw}{\textrm{crw}}
\newcommand{\qb}{{\bar{q}}}

\newcommand{\wb}{\bar{\omega}}
\newcommand{\wz}{\omega_0}
\newcommand{\bP}{{\bar P}}
\newcommand{\cP}{{{\overline {\cal P}}}}

\newcommand{\tC}{{\tilde C}}

\newcommand{\barLL}{{\bar {\mathcal{L}}}}
\newcommand{\class}{\textrm{cl}}

\newcommand{\sAAK}{{\scriptscriptstyle {\rm AAK}}}

\newcommand{\qqph}{\qquad \phantom{.}}
\newcommand{\qph}{\quad \phantom{.}}

\newcommand{\deco}{\textrm{dec}}
\newcommand{\tauphi}{{\tau_\varphi}}
\newcommand{\tauphiAAKone}{{\tau^\sAAK_{\varphi,1}}}
\newcommand{\gammaphiAAKone}{{\gamma^\sAAK_{\varphi,1}}}
\newcommand{\WL}{\textrm{WL}}
\newcommand{\cond}{\textrm{cond}}

\newcommand{\pbzero}{\textrm{pp}}

\newcommand{\tildet}{\widetilde t}

\newcommand{\Certt}{\tilde C^\ve ({\bmr_{12}}; t_1, t_2)}
\newcommand{\Coertt}{\tilde C^{(1)\ve} ({\bmr_{12}}; t_1, t_2)}
\newcommand{\Coerttph}{\tilde C^{(1)\ve}_\deco 
({\bmr_{12}}; t_1, t_2)}

\newcommand{\Fertt}{\tilde F^\ve ({\bmr_{12}}; t_1, t_2)}
\newcommand{\Foneertt}{\tilde F^{\ve} ({\bmr_{12}}; t_1, t_2)}

\newcommand{\Hikami}{\textrm{Hikami}}
\newcommand{\self}{\textrm{self}}
\newcommand{\selfvertex}{\textrm{s+v}}

\newcommand{\vertex}{\textrm{vert}}

\newcommand{\full}{\textrm{full}}
\newcommand{\bare}{\textrm{bare}}

\newcommand{\nud}{\nu} 
\newcommand{\Dd}{D} 

\newcommand{\GZ}{\textrm{GZ}}

\newcommand{\tauH}{{\tau_H}}

\newcommand{\gammaH}{{\gamma_H}}

\newcommand{\tauel}{{\tau_\textrm{el}}}
\newcommand{\lel}{{l_\textrm{el}}}
\newcommand{\kF}{k_\textrm{F}}

\newcommand{\Sec}[1]{Sec.~\ref{#1}}         
\newcommand{\Ref}[1]{Ref.~[\onlinecite{#1}]}  

\newcommand{\bmr}{{\bm{r}}}

\newcommand{\bmq}{{\bm{q}}}

\newcommand{\bnabla}{{\bm{\nabla}}}

\newcommand{\bcL}{\bar {\cal L}{}}

\newcommand{\ttau}{{\widetilde \tau}}

\newcommand{\bcG}{{\overline {\cal G}}}

\newcommand{\bcCEq}{{\overline {\cal C}{}^{\cal E}_\bmq}}

\newcommand{\bSigma}{{\overline \Sigma}{}}

\newcommand{\bcC}{{\overline {\cal C}}{}}
\newcommand{\bcD}{{\overline {\cal D}}{}}

\newcommand{\bLRbbqw}{{{\bcL^R_\bbmq (\bomega)}}}
\newcommand{\bLAbbqw}{{{\bcL^A_\bbmq (\bomega)}}}
\newcommand{\bLKbbqw}{{{\bcL^K_\bbmq (\bomega)}}}

\newcommand{\bomega}{{\bar \omega}}
\newcommand{\tomega}{{\widetilde \omega}}

\newcommand{\bbmq}{{\bar \bmq}}

\newcommand{\WK}{\widetilde W_{\class}}
\newcommand{\WKbare}{\widetilde W'_\class}
 \newcommand{\Eq}[1]{Eq.~(\ref{#1})} 
\newcommand{\Eqs}[1]{Eqs.~(\ref{#1})} 

 \newcommand{\EqI}[1]{Eq.~(I.\ref{#1})} 
 \newcommand{\EqsI}[1]{Eqs.~(I.\ref{#1})}

\begin{document}

\include{labelsI}
%
%

\title{Decoherence in weak localization II: 
Bethe-Salpeter calculation of Cooperon}


\author{
Jan von Delft,${}^1$ Florian Marquardt,${}^1$ 
R. A.   Smith,${}^2$ Vinay Ambegaokar${}^3$} \affiliation{$^1$ 
  Physics Department, Arnold Sommerfeld Center for Theoretical
  Physics, and Center for NanoScience,
  Ludwig-Maximilians-Universit\"at M\"unchen, 80333 M\"unchen,
  Germany\\
  $^2$School of Physics and Astronomy, University of Birmingham,
  Edgbaston, Birmingham B15 2TT, England\\
  $^3$Laboratory of Atomic and Solid State Physics Cornell University Ithaca, New York 14850, USA}

\date{October 21, 2005}

\begin{abstract} 

  This is the second in a series of two papers (I and II) on the
  problem of decoherence in weak localization. In paper I, we
  discussed how the Pauli principle could be incorporated into an
  influence functional approach for calculating the Cooperon
  propagator and the magnetoconductivity.  In the present paper II, we
  check and confirm the results so obtained by diagrammatically
  setting up a Bethe-Salpeter equation for the Cooperon, which
  includes self-energy and vertex terms on an equal footing and is
  free from both infrared and ultraviolet divergencies. We then
  approximately solve this Bethe-Salpeter equation by the Ansatz $\tC
  (t) = \tC^0 (t)e^{-F(t)}$, where the decay function $F(t)$
  determines the decoherence rate. We show that in order to obtain a
  divergence-free expression for the decay function $F(t)$, it is
  sufficient to calculate $\tilde C^1 (t)$, the Cooperon in the
  position-time representation to first order in the interaction.
  Paper II is independent of paper I and can be read without detailed
  knowledge of the latter.
\end{abstract}
\maketitle


\section{Introduction}

This is the second in a series of two papers (I and II), in which we
revisit the problem of decoherence in weak localization, using both an
influence functional approach (paper I) and a Bethe-Salpeter equation
for the Cooperon (paper II) to calculate the magnetoconductivity. The
basic challenge is to calculate the interference between two
time-reversed trajectories of an electron travelling diffusively
\emph{in a Fermi sea} and coupled to a noisy quantum environment, while taking
proper account of the Pauli principle.  In paper I \cite{paperI}, we
discussed how this could be done using an influence functional
approach by dressing the spectrum of the noise field by ``Pauli factors''
[see \EqI{subeq:theGreatReplacementa}; throughout, ``I'' will indicate
formulas from paper I]. Moreover, within the influence functional
scheme we concluded that a divergence-free calculation of the
decoherence rate can be obtained by expressing the Cooperon in the
position-time representation as
\begin{eqnarray}
  \label{eq:CfullC1preview}
\label{eq:CooperondecaywithF}
\label{completeAvgRRW}
  \tC(0,t)\simeq \tCO(0,t) \, 
  e^{-F(t)} \; ,
\quad 
F(t) = - \frac{\tC^{1}(0,t)}{\tCO(0,t)} 
 \; .  \label{Capprox1}
\end{eqnarray}
where $\tilde C^1 (0,t)$ is the first-order term in an expansion of
the full Cooperon $\tC (0,t)$ in powers of the interaction.  [In the
present paper, this statement will be made more precise: when
reexponentiating, a part of $\tilde C^1 (0,t)$ has to be omitted that
can be determined, in a self-energy-only calculation, to contribute
only to the prefactor of the Cooperon, see \Sec{sec:standardDyson}.]

These conclusions of paper I rested entirely on the influence
functional approach, and, in the discussion of the Pauli principle,
relied on heuristic arguments. Though these are in accord with results
derived elsewhere\cite{AAK,AAG,CS,Abrahams,vonDelft04} (as shown in Paper 1,
Section~\ref{app:GZrelation}), 
it is desirable to compare the approximations used 
and the results obtained so far against a treatment relying purely on
diagrammatic perturbation theory, the framework within which most of
our understanding of disordered metals to date has been obtained.

In the present paper II, we check and confirm the results mentioned
above by diagrammatically setting up a Bethe-Salpeter equation for the
Cooperon using standard Keldysh diagrammatic perturbation theory
(using conventions summarized in \cite{vonDelft04}), which includes
self-energy and vertex terms on an equal footing and is free from both
infrared and ultraviolet divergencies. We then show that this equation
can be solved (approximately, but with exponential accuracy) with an
Ansatz that is precisely of the form \Eq{Capprox1}, and that the
function $F(t)$ so obtained agrees with the form derived in paper I
[\EqI{exponentCLpaulimodified}].

  The usual diagrammatic calculation of the
Cooperon starts from a Dyson equation for a  
``self-energy-diagrams-only'' version of the Cooperon,
\begin{eqnarray}
\label{eq:standardDyson}  
\phantom{\bmq} 
{\bcC}_{{\ve}, \bmq}^{\self} (\omega) 
=
\bcC^0_\bmq (\omega) \left[ 1 + \bSigma^{\self}_{\ve , \bmq}
  (\omega)\, 
\bcC_{\ve, \bmq}^{\self} (\omega) 
\right] \; .
\end{eqnarray}
Here the Cooperon self-energy $\bSigma^{\self}_{\ve , \bmq}$
includes \emph{only self-energy diagrams}, in which interaction lines
connect only forward to forward or backward to backward electron
propagators; for these diagrams, the frequency labels along both the
forward and the backward propagators 
are \emph{conserved separately},
which is why the Dyson equation is a simple algebraic equation for
$\bcC_{\ve, \bmq}^{\self} (\omega)$.  However, the Cooperon
self-energy $\bSigma^{\self}_{\ve , \bmq} (\omega)$ turns out 
to be infrared
divergent in the quasi 2- and 1-dimensional cases. This problem  is usually
cured by inserting an infrared cutoff by hand (as reviewed in
Section~\ref{sec:standardDyson} below).  The results so obtained are
qualitatively correct but, due to the ad hoc treatment of the cutoff,
not very accurate quantitatively [\eg\ in the first line
of \Eq{eq:defineFeselfresult} for $\tilde F_\ve^\self (t)$ 
below the exponent is correct, but the prefactor is
wrong by roughly a factor of 2 compared to \EqI{eq:finalFrrw}]. 

Our goal in the present paper is to obtain, starting from a diagrammatic
equation, results free from any cutoffs, infrared or ultraviolet, that
have to be inserted by hand -- the theory ``should take care of its
divergencies itself''. This can be achieved if the Cooperon
self-energy is taken \emph{to include vertex diagrams}, in which
interaction lines connect forward and backward electron propagators. 
Since this brings about frequency transfers between the forward and
the backward propagators, their frequency labels are no longer
conserved separately.  As a consequence, it becomes necessary to study
a more complicated version of the Cooperon, $\bcC^{\cal E}_\bmq
(\Omega_1, \Omega_2) $, labeled by three frequencies, and governed not
by a simple algebraic Dyson equation, but by a nonlinear integral
equation, which we shall refer to as ``Bethe-Salpeter equation''. 

Finding an exact solution to the Bethe-Salpeter equation seems to be
an intractable problem, which we shall not attempt to attack. Instead,
we shall transcribe the Bethe-Salpeter equation from the
momentum-frequency to the position-time domain, in which it is easier
to make an informed guess for the expected behavior of the solution.
Using the intuition developed in paper I within the
influence-functional approach [summarized in \Eq{Capprox1} of the
present paper], we shall make an exponential Ansatz for
$\tC^\ve(r_{12},t_1,t_2)$, the Cooperon in the position-time domain.
We shall show that this Ansatz solves the Bethe-Salpeter equation with
exponential accuracy, in the sense that improving the Ansatz would
require terms to be added to the exponent that are parametrically
smaller (in powers of $1/g$, $g$ being the dimensionless conductance)
than the leading term in the exponent.


\begin{figure}
\begin{center}\includegraphics[%
  width=0.8\columnwidth]{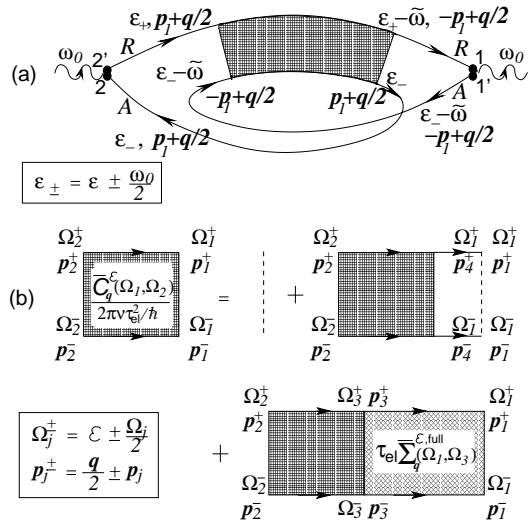}
\end{center}
\caption{\label{fig:conductivity} (a) Diagram for the weak
  localization correction to the AC conductivity, $\delta \sigma^\WL
  (\omega_0)$ [\Eqs{subeq:finalsigmawithintmaintextpert}].
  In contrast to the so-called
  ``interaction corrections'' to the conductivity, each current vertex
  is attached to both a retarded and advanced electron line.  (b)
  Diagrammatic definition of full Cooperon $ \bar {\cal C}_\bmq^{{\cal
      E}} (\Omega_1, \Omega_2)$, and schematic depiction of the
  Bethe-Salpeter equation (\ref{eq:Bethe-Salpeter-full-compact})
  satisfied by it; ${\cal E} = \toh (\Omega_j^+ + \Omega_j^-)$ is the
  conserved average of the energies of the upper and lower lines,
  while $\Omega_1$ and $\Omega_2$ are the outgoing and incoming
  Cooperon frequencies (with $\Omega_j = \Omega_j^+ - \Omega_j^-$).
  For the structure of the Cooperon self-energy $\bSigma^\full$ and
  details of our diagrammatic conventions, see Appendix~\ref{app:BS},
  Fig.~\ref{fig:bethesalpeter}.  }
\end{figure}

\section{Setting up Bethe-Salpeter Equation for Cooperon}

\subsection{Various expressions for conductivity}
\label{sec:conductivity-diagrams}

The diagrammatic definition of the weak localization contribution to
the AC conductivity of a quasi $d$-dimensional
disordered conductor is given by Fig.~\ref{fig:conductivity}(a), which
corresponds to the following expression (cf. App.~A of \Ref{AAK}, 
or App.~C of \Ref{vonDelft04}):
\begin{subequations}
\label{subeq:finalsigmawithintmaintextpert}
\begin{eqnarray}
  \label{eq:finalsigmawithintmaintextpert}
\label{eq:magnetoconductancefull}
& &  \delta \sigma^{\WL}_d (\wz)  =   - { \sigma_d  \over \pi \nud  \hbar} 
\left \langle \tilde {\cal C}_\cond^{\varepsilon, \wz}
\right \rangle_{\ve} \; ,  \qqph
\\
\label{eq:defineCcond}
& & \tilde {\cal C}_\cond^{\varepsilon, \wz} 
 =    \! \! 
 \int (d \, 2 \tomega) \!\! \int (d \bmq ) \, 
\overline {\cal C}^{\ve - {1 \over 2} \tomega}_\bmq
(\wz - \tomega ,  \wz + \tomega) 
 , \qqph  
\end{eqnarray}
\end{subequations}
where 
\begin{eqnarray}
\label{eq:average}
\bigl \langle ... \bigr \rangle_{\ve} \equiv
\int d \varepsilon 
\,  { f ( \vem ) - f ( \vep) \over \wz } \dots
\end{eqnarray}
denotes an average over $\ve$, with $\ve_\pm = \ve \pm \toh \wz$, 
and 
in the DC limit $\wz \to 0$ the weighting function
reduces to $ - f' (\ve)$, the
derivative of the Fermi function $f(\ve) = 1/[e^{\ve/T} +1]$. 
(In this paper, temperature is measured in units of frequency, \ie\
  $T$ stands for $k_{\rm B} T / \hbar$ throughout; 
 likewise, although $\ve$ will often be referred to as ``excitation
  energy'', it stands for a frequency.)

 The full Cooperon with general
arguments, $\overline {\cal C}^{\cal E}_\bmq (\Omega_1, \Omega_2)$ is
defined diagrammatically in Fig.~\ref{fig:conductivity}(b): ${\cal E}$
is the average of the frequencies of the upper and lower electron
lines, while $\Omega_1$ and $\Omega_2$ are the outgoing and incoming
Cooperon frequencies, respectively. In the absence of external
time-dependent fields, the average energy ${\cal E}$ is conserved
between incoming and outgoing lines.  
The Cooperon needed for the AC
conductivity in Fig.~\ref{fig:conductivity}(a) has incoming upper and
lower electron lines with energies $\vep$ and $\vem - \tomega$, 
and outgoing upper and lower electron lines with
energies $\vep  - \tomega$ and $\vem$,
implying $\Omega_1 = (\wz - \tomega)$, $\Omega_2 = ( \wz +
\tomega)$ and ${\cal E} = \ve - \toh \tomega$, as used in
\Eq{eq:defineCcond}.

To make contact with the expression for the conductivity in the
position-time representation used in paper I, we rewrite
\Eq{eq:defineCcond} as
\begin{subequations}
\begin{eqnarray}
\label{eq:Ctimeintegral}
\tilde {\cal C}_\cond^{\ve, \wz}   & \equiv &  
\int dt \, e^{i \wz t} \, 
\tilde C^{\vep}  (0 , t)
\\
\label{eq:C0tfromcond} 
\tilde C^\vep (0 ,  t) & \equiv & 
\tilde C^\vep (r_{12} = 0; t_1 = \toh t, t_2 = -\toh t) 
\; , \qqph 
\end{eqnarray}
where $\tilde C^\vep ({\bmr_{12}}; t_1, t_2)$
 is a representation of the full Cooperon in an
energy/position/two-time representation,
\begin{eqnarray}
\tilde C^\vep ({\bmr_{12}}; t_1, t_2) & = & 
\nonumber 
\int (d \bmq) (d \Omega_1) (d \Omega_2) \, 
 e^{i [ \bmq \bmr_{12}  - \Omega_1 t_1 + \Omega_2 t_2] } \,  
\\  \label{eq:Cetildewtaunew} 
& & \times \; 
   \bcC_\bmq^{\vep - {1 \over 2 } \Omega_2} (\Omega_1, \Omega_2) \; ,
  \label{eq:Cetildewtau}
\\
& = & 
\int (d \bmq) (d \, 2 \tomega) (d \omega) \,
 e^{i  [\bmq \bmr_{12} - \omega t_{12} + \tomega\tilde t_{12}]} \,  
\nonumber
\\
& & \label{eq:Cwtildew}
\times \;   
\bcC_\bmq^{\vep - {1 \over 2 } (\tomega+ \omega)} 
(\omega - \tomega, \omega + \tomega)  \; , 
\end{eqnarray}
\end{subequations}
where $r_{12} = r_1 - r_2$, $t_{12} = t_1 - t_2$, $\tilde t_{12} = t_1
+ t_2$, $\omega = \toh (\Omega_2 + \Omega_1)$ and $\tomega = \toh
(\Omega_2 - \Omega_1)$.  The $\int dt$ time integral in
\Eq{eq:Ctimeintegral} sets $\omega = \wz$ and hence $\vep - \toh
(\tomega + \omega) = \ve - \toh \tomega$ in \Eq{eq:Cetildewtau}, as
needed for \Eq{eq:defineCcond}. Inserting \Eq{eq:Ctimeintegral} into
(\ref{eq:finalsigmawithintmaintextpert}) we find
\begin{eqnarray}
\delta\sigma^\WL_d (\wz) & = & 
-\frac{\sigma_d}
{\pi\nu\hbar}\int_0^{\infty}dt\, 
e^{i \wz t} \, \left \langle \tC^{\vep} (0,t)
\right \rangle_{\ve} \, . \qqph 
\label{eq:IImagnetoconductance}
\end{eqnarray}
Thus, the DC limit $\delta \sigma^\WL_d (0)$ is seen to be an
energy-averaged version of \EqI{eq:magnetoconductance}.  Since our
goal is to make contact with the results of paper I, we shall take the
DC limit $\wz \to 0$ and $\vep \to \ve$ throughout below, (it is
straightforward to reinstate the $\wz$-dependence explicitly by
replacing the parameter $\ve$ by $\vep$ in all Cooperons below).

We choose to normalize the full Cooperons such that in
the absence of interactions, they reduce as follows to their
noninteracting versions ($\Omega_{12} = \Omega_1 - \Omega_2$):
\begin{subequations}
\begin{eqnarray}
  \label{eq:freefromfullCooperon}
\bar {\cal C}_\bmq^{{\cal E}} (\Omega_1, \Omega_2) 
& \stackrel{{\rm no \; int}}{\longrightarrow} &
2 \pi \, \delta (\Omega_{12}) \, \bcC^0_\bmq (\Omega_1) \; , 
\\
  \label{eq:freefromfullCooperon-time-domain}
\tilde C^{\cal E}  (\bmr_{12}; t_1, t_2) 
& \stackrel{{\rm no \; int}}{\longrightarrow} &
\tilde C^0 (\bmr_{12}, t_{12} ) \; . 
\end{eqnarray} 
\end{subequations}
Here $\bcC^0_\bmq (\Omega) = ( E_\bmq - i \Omega)^{-1}$, with $E_\bmq
= D \bmq^2 + \gammaH$, where $\gammaH$ is a magnetic-field induced
decay rate. For later reference, we also define 
$ E_\bbmq^0= D \bbmq^2$.

 Our strategy for determining the decoherence rate will be to find an
 approximation for $\tC^{\cal E}(0,t)$ of the form
 (\ref{eq:CfullC1preview}).  To this end, we shall set up a
 Bethe-Salpeter equation for $\bcC^{\cal E}_\bmq (\Omega_1, \Omega_2)
 $, transcribe it to the position-time domain to find a Bethe-Salpter
 equation for $\tC^{\cal E} (r_{12}; t_1, t_2)$, and then solve the
 latter using the Ansatz (\ref{eq:CfullC1preview}).

\subsection{Bethe-Salpeter Equation for Cooperon}
\label{sec:Bethe-Salpeter}

In the presence of interactions, the full Cooperon $\bar {\cal
  C}_\bmq^{{\cal E}} (\Omega_1, \Omega_2)$ satisfies a Bethe-Salpeter
equation of the  general form
\begin{eqnarray}
\label{eq:Bethe-Salpeter-full-compact}
\bcCEq (\Omega_1, \Omega_2) & = & 
\bcC^0_\bmq (\Omega_1) 
\left[ 2 \pi \, \delta (\Omega_{12}) \, 
\phantom{\int} \right. 
\\
\nonumber 
& & \left. + 
\int (d \Omega_3) \, 
\bSigma{}^{\cal E}_{\bmq, \full} (\Omega_1, \Omega_3) \, 
\bcCEq (\Omega_3, \Omega_2) \right] \; ,
\end{eqnarray} 
depicted schematically in Fig.~\ref{fig:conductivity}(b).  The average
energy ${\cal E}$ is conserved, because no external fields are
present. For the Cooperon self-energy $\bSigma{}^{\cal E}_{\bmq,
  \full} (\Omega_1, \Omega_3) $ occuring herein, we shall adopt the
diagrammatic definition first written down in Ref.~\onlinecite{AAV}.
The corresponding diagrams and equations for $\bSigma_\full$ are
rather unwieldy and hence have been relegated to Appendix~\ref{app:BS}
[see Fig.~\ref{fig:bethesalpeter} and \Eqs{subeq:newselfenergies} in
Appendix~\ref{app:BSdiagrams}]. This very technical appendix can be
skipped by casual readers; its contents are summarized in the next two
paragraphs, and to be able follow the developments of the main text
below, it should suffice to just occasionally consult the final
formulas for the self-energies given in \Eqs{subeq:bareselfenergies}.

The Cooperon self-energy $\bSigma{}_\full$ is itself proportional to
the Cooperon $\bcC$, thus the Bethe-Salpeter equation
(\ref{eq:Bethe-Salpeter-full-compact}) is non-linear in $\bcC$.
Solving it in full glory thus seems hardly feasible. Therefore, we
shall henceforth consider only a ``linearized'' version thereof,
obtained [in Appendix~\ref{sec:barSigma}] by replacing the full
Cooperon self-energy $\bSigma_\full$ in
\Eq{eq:Bethe-Salpeter-full-compact} by a bare one, $\bSigma_\bare$.
The latter, given explicitly in \Eqs{subeq:bareselfenergies}, is
obtained by making the replacement $\bcC^{\cal E }_{\bmq } (\Omega_1 ,
\Omega_3) \to
2 \pi \delta (\Omega_1 - \Omega_3) \, \bcC^0_\bmq (\Omega_1) $ for
every occurence of the full Cooperon in $\bSigma_\full$.

A perturbative expansion of the full Cooperon $\bcC$ in powers of the
interaction can readily be generated by iterating
\Eq{eq:Bethe-Salpeter-full-compact}. This is done explicitly to second
order in Appendix~\ref{sec:2ndorderexpansion} [see
\Eq{subeq:iterateBS}]. The expansion illustrates two important points:
firstly, due to the frequency transfers between the forward and
backward propagators generated by the vertex diagrams, the frequency
arguments of $\bSigma_\bare$ increasingly become ``entangled'' from
order to order in perturbation theory, \ie\ they occur in increasingly
complicated combinations. This makes it exceedingly difficult to
directly construct an explicit solution.  Secondly, no ultraviolet
divergencies arise in perturbation theory, confirming the heuristic
Golden Rule arguments of Paper I, Section~\ref{sec:goldenrule} (and
contradicting suggestions to the contrary implicit in
Refs.~\onlinecite{GZ2,GZ3,GZ7}; see Appendix~\ref{sec:2ndorderexpansion}
for a discussion of this point).

\subsection{Recover Dyson Equation by Neglecting Vertex Terms}
\label{sec:standardDyson}

Before attempting to solve the (linearized) Bethe-Salpeter equation,
it is instructive to start for the moment with a rather strong
approximation, namely to simply neglect all vertex terms (they will be
reinstated later), thereby avoiding the abovementioned
``entanglement'' of frequencies.  This reduces the Bethe-Salpeter
equation to the more familiar Dyson equation (\ref{eq:standardDyson})
for the ``self-energy-only'' Cooperon  $\bcC^\self$,
and will allow us to review some standard arguments and to recover
some familiar and simple results. 

In the absence of vertex diagrams, the Cooperon self-energy $\bSigma{}^{\cal
  E}_{q, \bare} (\Omega_1, \Omega_2)$ of
\Eq{eq:Bethe-Salpeter-new-selfenergy} is proportional to $\delta
(\Omega_{12})$, implying the same for the Cooperon $ \bcC_\bmq^{\ve -
  {1 \over 2 } \Omega_2, \self} (\Omega_1, \Omega_2)$, so that the
Cooperons needed on the right-hand sides of \Eqs{eq:Cetildewtaunew}
and (\ref{eq:C0tfromcond}) can, respectively, be written as
\begin{subequations}
  \label{subeq:Cooperonwithoutvertex}
\begin{eqnarray}
  \label{eq:Cooperonwithoutvertexw}
  \bcC_\bmq^{\ve - {1 \over 2 } \Omega_2, \self} (\Omega_1, \Omega_2)
  & \equiv  &
  2 \pi \delta (\Omega_{12}) \, 
  \bcC_{\ve, \bmq}^{\self} (\Omega_2) \; ,
\\
\tilde C^{\ve, \self} (0 , t)  & = &  
  \label{eq:Cooperonwithoutvertext}
 \int (d \bmq) (d \omega) \, e^{-i \omega t}
 \bcC_{\ve, \bmq}^{\self} (\omega)  \; . \qqph
\end{eqnarray}
\end{subequations}
The ``single-frequency'' Cooperon $\bcC_{\ve, \bmq}^{\self} (\omega) $
introduced in \Eq{eq:Cooperonwithoutvertexw} is the generalization of
the free, single-frequency Cooperon $ \bcC^0_\bmq (\omega)$ to the
case of a Cooperon for pairs of paths with average energy $\ve$ in the
presence of self-energy-only interactions.  From
\Eq{eq:Bethe-Salpeter-full-compact}, it is seen to satisfy the
familiar Dyson equation (\ref{eq:standardDyson}), with solution
\begin{eqnarray}
  \label{eq:Csolvednovertex}
\bcC_{\ve, \bmq}^{\self} (\omega) & = & {1 \over E_\bmq - i \omega 
- \bSigma{}^{\self}_{\ve, \bmq} (\omega) } \; , 
\end{eqnarray}
where the effective Cooperon self-energy
is given by an integral over all momentum and frequency transfers to
the environment,
\begin{eqnarray}
\label{eq:effectiveCooperonselfenergy}
\bSigma{}^{\self}_{\ve , \bmq}  (\omega)
& = & 
{1 \over \hbar} \int (d \bomega) (d \bbmq) \, 
\bSigma{}^{\ve , \self}_{\bmq, \bbmq, \bare} 
(\omega , \bomega) \, ,
\end{eqnarray}
with $\bSigma{}^{\ve , \self}_{\bmq, \bbmq, \bare} (\omega , \bomega) $
given in \Eqs{eq:selfenergy-selfI-new-old} to (\ref{eq:SigmabareZ}).

Now, the standard way to extract the decoherence rate from
\Eq{eq:Csolvednovertex} is to expand the self-energy in powers of
$E_\bmq - i \omega$:
\begin{subequations}
  \label{subeq:expandselfenergy}
\begin{eqnarray}
\label{eq:expandselfenergy}
\bSigma{}^{\self}_{\ve , \bmq}  (\omega) & = & 
- \gamma_{\ve,\bmq}^{\varphi, \self}  +
(E_\bmq - i \omega) \bSigma{}^{\prime \self}_{\ve, \bmq}  + \dots \; , 
\qqph
\\
\label{eq:definegammaphi}
\gamma_{\ve, \bmq}^{\varphi,\self} & \equiv & 
- \left[ \bSigma{}^{\self}_{\ve , \bmq}  (\omega) \right]_{E_\bmq =  i
  \omega  } \, . 
 \label{eq:definegammaphiagain}
\end{eqnarray}
\end{subequations}
The leading ``Cooperon mass'' 
term can be identified with the decoherence rate, because
\Eq{eq:Cooperonwithoutvertext} yields (after performing 
the  $\int (d \omega)$ integral by contour integration)
\begin{eqnarray}
  \label{eq:Cooperonselftimedomain}
  \tilde C^{\self}_\ve (0,t)
& \simeq &  
  \label{eq:Cooperonwithoutvertextexplicit}
 \! \! \int (d \bmq) \, e^{- t( E_\bmq + \gamma^{\varphi, \self}_{\ve, \bmq})}
\Bigl(1 + \bSigma{}^{\prime \self}_{\ve, \bmq} + \dots \Bigr)  . \qqph 
\end{eqnarray}
Since the $(d \bmq)$ integral is dominated by small $q$, 
let us replace $\bmq$ by 0 
in  $\gamma^{\varphi, \self}_{\ve, \bmq}$
and $\bSigma{}^{\prime \self}_{\ve, \bmq} $, so that
they can be pulled out of the integral. This yields 
\begin{subequations}
  \label{subeq:Cooperonwithoutvertextexplicitfinal}
\begin{eqnarray}
  \label{eq:Cooperonwithoutvertextexplicitfinal}
   \tilde C^{\self}_\ve (0,t) & \simeq & \tilde C^0 (0, t) \, 
e^{- \tilde F^\self_\ve
   (t)} \Bigl(1 + \bSigma{}^{\prime \self}_{\ve, 0} + \dots \Bigr) \; ,
\qqph \\
\label{eq:defineFeself}
\tilde F^\self_\ve (t) & = & t \, \gamma^{\varphi, \self}_{\ve, 0}  \; , 
\end{eqnarray}
\end{subequations}
in which $\tilde C^{\self}_\ve (0,t)$ is expressed in a form
reminiscent of \Eq{Capprox1}: a free Cooperon, times the exponential
of a decay function, times a factor $1 + \bSigma{}^{\prime
  \self}_{\ve, 0}$ that renormalizes the overall amplitude of the
Cooperon (\ie\ it corresponds to ``wave-function'' renormalization, in
analogy to the occurence of a finite quasiparticle weight $Z$ in a
Fermi liquid due to the short-time decay that is not resolved further
by this approximation).

Since we have to set $E_\bmq = i \omega$ in \Eq{eq:definegammaphi} and
$\bmq = 0$ in \Eq{eq:defineFeself}, it is natural to split the self
energy of \Eq{eq:effectiveCooperonselfenergy} into two parts,
$\bSigma{}^{\self}_{\ve, \bmq} (\omega) = \bSigma{}^{\self,
  \deco}_{\ve, \bmq} (\omega) + \bSigma{}^{\self, Z}_{\ve, \bmq}
(\omega) $, chosen such that $\bSigma{}^{\self,Z}_{\ve, \bmq} (\omega)
$ vanishes when $E_q = i\omega$ and $\bmq = 0$.  [This requirement is
in fact fulfilled by (and was the motivation for) the separation of
\Eq{eq:selfenergy-selfI-new-old} into two terms, labeled ``\deco'' and
``$Z$''.]  Thus, $ \gamma^{\varphi, \self}_{\ve,0}$ depends only on
$\bSigma{}^{\self, \deco}_{\ve, \bmq} (\omega) $; using
\Eq{eq:Sigmabaredeco} in \Eqs{eq:effectiveCooperonselfenergy}
and (\ref{eq:define:Ldeco}), it can
be written as follows for not too large magnetic fields\cite{gammaH/T}
($\gammaH / T \ll 1$):
  \begin{eqnarray}
    \label{eq:gammaselfdefine}
  \gamma^{\varphi, \self}_{\ve, 0}  & = &
\oneoverhbarsq \int (d \bomega) (d \bbmq) \, 
{2 E_\bbmq^0 
\over (E_\bbmq^0)^2 + \bomega^2} \, \bigl\langle V V  
  \bigr\rangle^\pbzero_{\qb \wb}  \; ,  \qqph
\label{eq:VeffNyquistcothtanh}
  \end{eqnarray}
Here the effective propagator 
$\bigl\langle \hat{V} \hat{V}   \bigr\rangle^\pbzero_{\qb \wb} $ 
arising in \Eq{eq:VeffNyquistcothtanh}
turns out to be precisely the Pauli-principle-modified propagator of
\EqI{theGreatReplacement} which we conjectured by heuristic arguments
in Paper I, Section~\ref{sec:symmPauli-mod}:
\begin{eqnarray}
\label{subeq:IItheGreatReplacementa}
\smalloneoverhbar  \bigl\langle V V  
 \bigr\rangle^\pbzero_{\qb \wb} 
\label{IItheGreatReplacementa}
& =  & 
\;
\textrm{Im} \bar  {\mathcal{L}}_{\qb}^{R}(\wb)
\left\{
  \coth \Bigl[{ \bomega \over 2T}\Bigr] \, \right.
\\  \nonumber 
& & 
 \left. + \toh 
\tanh\Bigl[ {\ve - \wb \over  2T} \Bigr] - 
\toh \tanh\Bigl[ {\ve + \wb \over  2T} \Bigr] 
 \right\} ,  
\label{eq:cancelspontabs}
\end{eqnarray}
  The $\coth + \tanh$
combination occuring in $\bigl\langle \hat{V} \hat{V}
\bigr\rangle^\pbzero_{\qb \wb} $ limits the frequency integral to
$| \bomega | \lesssim T$, as anticipated by the Golden Rule
discussion in Section~\ref{subsec:goldenrule} of paper I
[cf. \EqI{eq:cosh+tanh}].

After performing the $(d \bbmq)$ integral in \Eq{eq:gammaselfdefine},
the remaining $(d \bomega)$ integral turns out to have an infrared
divergence for quasi 1- or 2-dimensional samples.  To be explicit, if
we regularize it by hand by inserting a steplike cutoff function
$\theta (|\bomega| - \bomega_0)$, we obtain, for the
quasi-$d$-dimensional case
\begin{subequations}
  \begin{eqnarray}
\label{eq:gamma0selffirstexp}
    \gamma^{\varphi, \self}_{\ve, 0} 
& \simeq  & 
{p_d \over 2}
\!\! \int_{\bomega_0}^{\infty} \!\! 
{ d \bomega  \over \bomega^{1-d/2} }
\left\{ \coth \Bigl[{\vhbar \bomega \over 2T} \Bigr] \right.
\\ \nonumber 
& & \left. 
 + \toh \tanh \Bigl[{\varepsilon  \!-   \bomega \over  2T} \Bigr]
 - \toh \tanh \Bigl[{\varepsilon  \! +   \bomega \over  2T}\Bigr] 
\right\} \qqph  
\end{eqnarray}
with $p_1=\sqrt{2\gamma_1}/\pi$, $p_2=1/(2\pi g_2)$ and
$p_3=1/(\sqrt{2 \gamma_3} \pi^2)$, where $\gamma_1 = \Dd (e^2 / \hbar
\sigma_1)^2$, $g_2=\hbar \sigma_2/e^2$, and $\gamma_3 = \Dd (e^2 /
\hbar \sigma_3)^{-2}$. For $d=3$, the integral is well-behaved in the
limit $\wb_0 \to 0$, but not for $d=1,2$.  For example, in the quasi
1-dimensional case, the integral evaluates to
\begin{eqnarray}
    \gamma^{\varphi, \self}_{0, 0} 
& = & 
{2 T  \over   \pi } 
\left[{2 \gamma_1 \over \bomega_0} \right]^{1/2}
\left[ 1 + {\cal O} \left(
    \Bigl[ {\hbar \bomega_0  \over  T }\Bigr]^{1/2} \right) \right] , \qqph 
\label{eq:1d-divergence} 
\end{eqnarray}
\end{subequations}
which diverges for $\bomega_0 \to 0$.  This infrared divergence arises
because in the present approach we have neglected vertex terms, which
in general ensure that frequency transfers smaller than the inverse
propagation time $1/t$ do not contribute [cf.\ Paper I,
Section~\ref{sec:Cooperon-decay-for}]. Thus, we should choose the
infrared cutoff at $\bomega_0 \simeq 1/t$ (as noted in
Ref.~\onlinecite{vonDelft04}), obtaining a
\emph{time-dependent}\cite{Montambaux04} decay rate, $
\gamma^{\varphi, \self}_{\ve, 0} = 2 \sqrt{2 / \pi}\: t^{1/2}
/(\tauphiAAKone)^{3/2}$, where\cite{factorof2}
\begin{eqnarray}
\label {eq:IItauhphiAAK}
{1 \over \tauphiAAKone} =   \gammaphiAAKone =   
\left ( {T \sqrt \gamma_1 }\right)^{2/3}
= \left({ T e^2  \sqrt D \over \hbar \sigma_1} \right)^{2/3}  \; . \qqph
\end{eqnarray}
is the decoherence rate first derived by AAK \cite{AAK}.
$\gamma^{\varphi, \self}_{\ve, 0}$ 
grows with time, because with increasing time, the
Cooperon becomes sensitive to more and more modes of the interaction
propagator with increasingly smaller frequencies, whose contribution
in \Eq{eq:gamma0selffirstexp} scales like $\bomega^{-3/2}$.
 
Alternatively, instead of $\bomega_0 = 1/t$, the choice $\bomega_0 =
\gamma^{\varphi, \self}_{\ve, 0}$ is often made, since in weak
localization the time duration of relevant trajectories is set by the
inverse decoherence rate. Then \Eq{eq:1d-divergence} is solved
selfconsistently \cite{AAG,GZ2}, yielding $ \gamma^{\varphi,
  \self}_{\ve, 0} = (2 \sqrt 2 / \pi)^{2/3}/ \tauphiAAKone $, with
$\tauphiAAKone $ again given by \Eq{eq:IItauhphiAAK}. 

The decay functions for $d=1$ corresponding to the above two choices
of $\bomega_0$ in \Eq{eq:1d-divergence} are, respectively [from
\Eq{eq:defineFeself}]:
\begin{eqnarray}
\label{eq:defineFeselfresult}
\tilde F^\self_\ve (t) & = & \left\{ 
\begin{array}{c}
(2 \sqrt 2 / \pi) \, (t / \tauphiAAKone)^{3/2} \; , \rule[-3mm]{0mm}{0mm}
\\
(2 \sqrt 2 / \pi)^{3/2} \, (t / \tauphiAAKone) \; . 
\end{array}
\right. 
\end{eqnarray}
Evidently, both equations describe decay on the same time scale
$\tauphiAAKone$.  The second choice does not properly reproduce the
3/2 power law in the exponent that we expect from \EqI{eq:finalFrrw}
for $F^\class_\crw (t)$ (a fact strongly criticized by Golubev and
Zaikin (GZ) \cite{GZ3}).
However, the first choice does, up to a numerical prefactor, whose
precise value can not be expected to come out correctly here, because
it depends on the shape of the infrared cutoff function (arbitrarily
chosen to be a sharp step function above).  \emph{We thus recover the
  classical result for the decoherence rate}. The reason is
essentially that Pauli blocking (represented by the $\tanh$ terms in
$\bigl\langle V V \bigr\rangle^\pbzero_{\qb \wb}$) suppresses the
effects of quantum fluctuations (represented by the $+1$ in $\coth
(\wb/2T) = 2 n (\wb) + 1$) with frequencies larger than $T$, as
discussed in detail in Sec.~\ref{sec:symmPauli-mod} of paper I.
Moreover, we also obtain the important result that the first ``quantum
correction'' to this classical result that arises from self-energy
terms [the ${\cal O} \left([{\bomega_0 / T }]^{1/2}\right)$
correction in \Eq{eq:1d-divergence}], is smaller
by a factor  $\hbar /\sqrt(tT)$.
For $t \sim \tauphi$, this is $\ll 1$ in the regime of weak
localization [cf.\ discussion after
\EqI{eq:dimensionlessconductance}], in agreement with the conclusions
of Vavilov and Ambegaokar.\cite{AV}

\section{Bethe-Salpeter equation in the position-time domain}
\label{sec:solutionBS}

The infrared divergencies mentioned above are cured as soon as vertex
diagrams are included. However, as mentioned at the end of
Section~\ref{sec:Bethe-Salpeter} and detailed in
Appendix~\ref{sec:barSigma}, frequency-entanglement then renders the
momentum-frequency version (\ref{eq:Bethe-Salpeter-full-compact}) of
the Bethe-Salpeter equation intractable.  This suggests that we try a
more pragmatic way of finding an approximate expression for the full
Cooperon: inspired by the insight from Paper~I
(Sec.~\ref{eq:generalF}) that in the case of classical noise, a rather
accurate description of the Cooperon can be obtained \emph{in the
  position-time representation} by reexponentiating its expansion to
first order in the interaction [\Eq{Capprox1} of paper II], we shall
try a similar approach here: we transcribe the Bethe-Salpeter equation
to the position-time domain to obtain an equation for the
corresponding Cooperon $\tilde C{}^\ve (r_{12}, t_1, t_2)$ of
\Eq{eq:Cetildewtaunew}, and solve this equation approximately with an
exponential Ansatz; this Ansatz will turn out to yield precisely the
reexponentiation of $\tilde C^{\ve (1)} (r_{12}; t_1,_2t)$, the
first-order expansion of the full Cooperon, in full analogy to
\Eq{Capprox1}.

\subsection{Transcription to time domain, exponential Ansatz}
  
Let us now consider the Bethe-Salpeter
\Eq{eq:Bethe-Salpeter-full-compact} for $ \bcC_\bmq^{\ve - {1 \over 2
  } \Omega_2} (\Omega_1, \Omega_2)$, \ie\ with ${\cal E} = \ve - \toh
\Omega_2$, as needed in \Eq{eq:defineCcond} when $\wz = 0$.  This
equation can be transcribed, using \Eq{eq:Cetildewtau} (with $\vep$
there replaced by $\ve $) to the form
\begin{eqnarray}
\nonumber
 &  & \bigl( - D \nabla^2_{\bmr_1} + \partial_{t_1} +  \gammaH \bigr) \, 
\Certt
 =  
\delta (\bmr_{12}) \, \delta (t_{12})  + 
\\
\label{subeq:Cetildewtau}  
  \label{eq:BSpositiontau}
& & 
\int d \bmr_4 \, d t_4 \, d t^\prime_4 \;
 \tilde \Sigma^{\ve, t^\prime_4}_\full ({\bmr_{14}} ; t_1,t_4) \,  
\tilde C^\ve (\bmr_{42}; t_{44'}, t_{24'}) \; ,  
\end{eqnarray}
where the self-energy in the energy/position/times representation
is defined by:
\begin{eqnarray}
\label{eq:CSigmaetildewtaut4prime}
 \tilde \Sigma^{\ve ,t^\prime_4}_\full ({\bmr_{14}} ; t_1,t_4) \,  
 & \equiv & 
\int (d \bmq) (d \Omega_1) (d \Omega_2) (d \Omega_4) \, 
\\
\nonumber
& & \phantom{.} \hspace{-2.5cm} 
\times e^{i [\bmq \bmr_{14} -  \Omega_1 t_1 + \Omega_4 t_4  
- t^\prime_4 (\Omega_2  - \Omega_4) ]}
\, \bSigma{}_{\bmq, \full}^{\ve - {1 \over 2 } \Omega_2} 
(\Omega_1,  \Omega_4)  \; . 
\end{eqnarray}

Before trying to solve \Eq{eq:BSpositiontau}, let us get a feeling for
the structure of this equation, by calculating the zeroth and first
order terms of $\Certt$ in an expansion in powers of the interaction
propagator (\ie\ $\tilde \Sigma_\bare$). To this end, we use the fact
that 
\begin{eqnarray}
 \bigl( - D \nabla^2_\bmr + \partial_t + \gammaH \bigr)\, \tilde C^0
(\bmr, t) = \delta (\bmr) \, \delta (t) \; , 
\end{eqnarray} 
iterate \Eq{eq:BSpositiontau} once, and replace $\tilde \Sigma_\full$
by $\tilde \Sigma_\bare$ [given by \Eqs{subeq:bareselfenergies}] on
its right-hand side. We find
\begin{subequations}
  \label{subeq:Crtau01}
  \begin{eqnarray}
\nonumber
\Certt & = & 
 \tilde C^0 (\bmr_{12} , t_{12})  +  \Coertt +  \dots \; , 
\\
    \label{eq:Crtaufull}
\end{eqnarray}
where $\Coertt$ has just the structure discussed in Paper I,
 Sec.~\ref{sec:rrw-vs-urw}, describing propagation from
$(r_2,t_2)$ to $(r_1,t_1) $, with interaction vertices along the way
at points $(r_4, t_4)$ and $(r_3, t_3)$:
\begin{eqnarray}
\Coertt
& = & 
\int d \bmr_3 \, dt_3 \, d \bmr_4 \, d t_4 \:
\tilde C^0 (r_{13}, t_{13}) 
\label{eq:C1newdefine}
      \label{eq:Crtau1partial}      \label{eq:Crtau1}
\\ \nonumber
& & \times 
\tilde \Sigma^\ve_\bare (\bmr_{34} ; t_3,t_4) \,  
\tilde C^0 (r_{42}, t_{42}) \; ,
\\
 \tilde \Sigma^\ve_\bare ({\bmr_{34}} ; t_3,t_4) \,  
& \equiv & 
\int d t^\prime_{4} \;  \tilde \Sigma^{\ve, t^\prime_4}_\bare
({\bmr_{34}} ; t_3,t_4) 
\\
 & = & 
\int (d \bmq) (d \Omega_3) (d \Omega_4) \, 
e^{i [\bmq \bmr_{34} - \Omega_1 t_3 + \Omega_4 t_4] } \,  
\nonumber 
\\
\label{eq:CSigmaetildewtau}
& & 
\times 
\bSigma{}_{\bmq, \bare}^{\ve - {1 \over 2 } \Omega_4} 
(\Omega_3,  \Omega_4) \; . 
  \end{eqnarray}
\end{subequations}

Let us  now construct an approximate solution of the Bethe-Salpeter
equation (\ref{eq:BSpositiontau}), by making an exponential
Ansatz of the following form:
\begin{eqnarray}
  \label{eq:AnsatzCewtau}
\Certt & = & 
 \tilde C^0 (\bmr_{12} , t_{12})  \, 
e^{- \Fertt } \; . 
\end{eqnarray}
The decay function $\tilde F^\ve$ is needed only for $t_{12} \ge 0$
(since $ \tilde C^0 (\bmr_{12} , t_{12})$ vanishes otherwise), and is
required to obey the initial condition $\tilde F^{\ve}(0, t_2, t_2) =
0 $ for all $t_2$.  The Ansatz (\ref{eq:AnsatzCewtau}) solves
\Eq{eq:BSpositiontau} exactly, provided that the decay function
$\tilde F^\ve$ satisfies the equation
\begin{widetext}
\begin{eqnarray}
\nonumber
- \lefteqn{  \tilde C^0 (\bmr_{12} , t_{12}) \left\{ 
\left[
 \partial_{t_1} -  D \nabla^2_{\bmr_1} - 2 D
 {\nabla_{\bmr_1} \tilde C^0 (\bmr_{12} , t_{12})  
\cdot \bnabla_{\bmr_1} \over 
\tilde C^0 (\bmr_{12} , t_{12}) } \right] \Fertt 
- D \bigl[\nabla_{\bmr_1} \Fertt \bigr]^2 \right\} } \qqph \qqph
\\ 
& = & 
  \label{eq:eqforFfunction}
\label{subeq:BSforfullcooperon}
\int d \bmr_4 \, d t_4 \, d t^\prime_4 \; 
 \tilde \Sigma^{\ve, t^\prime_4}_\full (\bmr_{14} ; t_1,t_4) \,  
\tilde C^0 (\bmr_{42}, t_{42}) \, 
e^{- [\tilde F^{\ve} (\bmr_{42}; t_{44'}, t_{24'})   - 
\tilde F^{\ve} (\bmr_{12}; t_1, t_2)] } \; . 
\end{eqnarray}
\end{widetext}

 \subsection{Evaluation of the decay function $\tilde F^\ve (t)$}
\label{sec:evaluatingFe}

Let us now evaluate the decay function $\tilde F^\ve (t)$ explicitly;
after three simplifying approximations, we shall find that it
reproduces the function $F^\pp_{d,\crw} (t)$ of
\EqI{exponentCLpaulimodified}.

Our first simplifying approximation is as follows: instead of trying
to solve \Eq{eq:eqforFfunction} in general, we shall be content to
\emph{determine the decay function $\tilde F^\ve$ only to linear order
  in the self-energy,} in accord with the fact that we 
``linearize'' the latter by replacing the full self energy by the bare
one. (Including nonlinear contributions would add terms that are
smaller than those kept by powers of the small parameter $1/g$.)
To this end, it suffices to
linearize \Eq{eq:eqforFfunction} in $\tilde F^\ve$, by dropping the $
(\nabla_{\bmr_1} \tilde F^\ve)^2 $ term on the left-hand side and the
exponential factor $e^{-[\tilde F^\ve_{42}-\tilde F^\ve _{12}]}$ on
the right-hand side, and replacing $\tilde \Sigma_\full$ by $\tilde
\Sigma_\bare$.  One readily finds that the resulting linearized
equation is solved by
\begin{eqnarray}
  \label{subeq:linearizedFequation}
  \label{eq:linearizedFequation}
\Foneertt & = & - \;  {\Coertt  \over  \tilde C^0 (\bmr_{12} , t_{12}) } \; , 
\end{eqnarray}
where $\tilde C^{(1)\ve}$ is given by \Eq{eq:Crtau1}.  Thus, the
expansion of $\tilde C^\ve $ [\Eq{eq:AnsatzCewtau}] to first order in
$\tilde F^\ve$ reproduces \Eq{eq:Crtaufull}, as it should, and conversely,
\Eq{eq:AnsatzCewtau} turns out to be nothing but the reexponentiated
version of \Eq{eq:Crtaufull}.  Our explicit solution of the
Bethe-Salpeter equation, to linear-in-$\tilde \Sigma$ accuracy in the
exponent, thus very nicely confirms the heuristic analysis presented
in Section~\ref{eq:generalF} of Paper I in favor of reexponentiation
strategies.

The second approximation is necessitated by the first: upon comparing
with the structure of the self-energy-only solution
[\Eq{eq:Cooperonwithoutvertextexplicitfinal}], and following the
discussion before \Eq{eq:gammaselfdefine}, we recognize that
effectively only a part of $C^{(1)\ve}$ may be reexponentiated (note
that this remark would be irrelevant if we were able to find the
\emph{exact} $\tilde F^\ve$). Therefore, when evaluating $\tilde
C^{(1)\ve}$ explicitly from \Eq{eq:C1newdefine}, we insert
$\bSigma_\bare = \bSigma_\bare^\self + \bSigma_\bare^\vertex$
[\Eqs{eq:Bethe-Salpeter-new-selfenergy}] into
\Eq{eq:CSigmaetildewtau}, but for the self-energy term
$\bSigma_\bare^\self$ [\Eq{eq:selfenergy-selfI-new-old}] we retain
only the ``decoherence'' contribution $\bSigma^{\self, \deco}_\bare$
[\Eq{eq:Sigmabaredeco}], because $\bSigma^{\self, Z}_\bare$
[\Eq{eq:SigmabareZ}] contributes only to the renormalization of the
overall amplitude of the Cooperon (and in any case its time-dependence
for long times turns out to be weaker than that arising from
$\bSigma^{\self, \deco}_\bare$, as is checked explicitly in
Appendix~\ref{sec:SigmaZsubleading}).  In other words, we write
$\tilde C^{(1),\ve} = \tilde C^{(1),\ve}_{\deco} + \tilde
C^{(1),\ve}_{\self,Z}$, and drop the second term.  The resulting
expression for $\tilde C^{(1),\ve}_{\deco} $ reads
\begin{eqnarray}
\nonumber
& & \phantom{.} \hspace{-0.5cm} 
\Coerttph 
  =
  { 1 \over \hbar} \int (d \bmq)  (d \omega)  
 (d \bbmq)  (d \bomega) \,  e^{i \bmq \bmr_{12}} 
\, e^{-i \omega t_{12}} 
\\
  \label{eq:C1tauexplicit}
& & \times
\Biggl\{  \bcC^0_\bmq (\omega ) \,
\bSigma{}^{\ve , \self, \deco}_{\bmq, \bbmq, \bare} 
(\omega , \bomega) \, 
  \bcC^0_\bmq (\omega ) \, 
\Biggr. 
\\ \nonumber
& & 
\quad 
  \Biggl. 
+ \, e^{ i  \bomega \tildet_{12}}
\,  \bcC^0_\bmq (\omega - \bomega) \,
\bSigma{}^{\ve , \vertex}_{\bmq, \bbmq, \bare} 
(\omega , \bomega) \,    \bcC^0_\bmq (\omega + \bomega) 
\Biggr\} \; . 
\end{eqnarray}
[The quickest way to arrive at \Eq{eq:C1tauexplicit} is from
the second term of \Eq{eq:Bethe-Salpeter-full-compact},
with $\bcCEq (\Omega_3, \Omega_2) \to 2 \pi \delta(\Omega_{23})
\bcC^0_q (\Omega_2)$ on the right-hand side, and 
$\bSigma_\full \to 
\bSigma_\bare $, given by \Eqs{subeq:bareselfenergies}.] 

Our third approximation for evaluating the first order decoherence
correction to the Cooperon (and thus the decay function) consists in
retaining only its dominating long-time behavior, for $T t_{12} \gg 1
$. In this limit, terms of order $\omega/T$ are $\ll 1$ and may be
neglected (they produce subleading contributions for $\tilde F^\ve
(t)$, as is checked explicitly in App.~\ref{app:checkF1}). This allows
us to keep only the $\omega = 0$ component of the effective
environmental propagator $\bcL^{\deco}_{{\cal E} \omega, \bbmq}
(\bomega) $ [\Eq{eq:define:Ldeco}], which is contained in both
$\bSigma{}^{\ve , \self, \deco}_{\bmq, \bbmq, \bare} $ and $
\bSigma{}^{\ve , \vertex}_{\bmq, \bbmq, \bare} $, see
\Eqs{eq:Sigmabaredeco} and (\ref{eq:selfenergy-vertI-new-old}).  More
formally, after substituting the latter two equations
for the $\bSigma{}$'s occuring in
\Eq{eq:C1tauexplicit} and  symmetrizing the integrand w.r.t.\ $\bomega
\leftrightarrow - \bomega$, we Taylor-expand $\bar {\cal
  L}^{\deco}_{\ve \omega, \bbmq} (\bomega)$ in powers of $\omega$
and represent $\omega$ as $ i \partial_{t_{12}}$ under the
Fourier integral, thereby bringing 
\Eq{eq:C1tauexplicit} into the form
\begin{eqnarray}
\nonumber
  \Coerttph 
&  = & -
\sum_{n = 0}^\infty \partial^n_{t_{12}}
\oneoverhbar \int (d \bbmq)  (d \bomega) \, 
 \bcL^{\deco}_{\ve (n), \bbmq} (\bomega) \, 
\\   \label{eq:Fonefirstresults}
& & \times  \, 
\tC^0 (r_{12}, t_{12}) \, \cP^\crw_{(r_{12}; t_1, t_2)} (\bbmq, \bomega) \; .
\qqph
\end{eqnarray}
Its ingredients are defined as follows:
\begin{eqnarray}
\bcL^{\deco}_{\ve (n), \bbmq} (\bomega) 
& = &  { (i \partial_\omega)^n  \over 2  n!} 
\Bigl[
  \bcL^{\deco}_{\ve \omega, \bbmq} (\bomega)
+  \bcL^{\deco}_{\ve \omega, \bbmq} (-\bomega)
\Bigr]_{\omega = 0} 
, \qqph
\label{eq:define:Ldeco:n}
\end{eqnarray}
\begin{widetext}
\begin{subequations}
\label{subeq:definePfull}
\begin{eqnarray}
\label{eq:Plongdefine}
\cP^\crw_{(r_{12}; t_1, t_2)} (\bbmq, \bomega) & = &
2 \int (d \omega)  (d \bmq) \, 
{e^{i \bmq \bmr_{12}}  \, e^{-i \omega t_{12}}  
\over \tilde C^0 (\bmr_{12} , t_{12})}
\left[\bcC^0_\bmq (\omega ) \, 
\bcC^0_{\bmq - \bbmq} (\omega - \bomega)  \, 
\bcC^0_\bmq (\omega )
- e^{i \bomega \tildet_{12}} \, 
\bcC^0_\bmq (\omega - \bomega ) \, 
\bcC^0_{\bmq - \bbmq} (\omega ) \,
\bcC^0_\bmq (\omega + \bomega) \, 
\right] \qquad 
\\
& = & 
\label{eq:freq-to-time-result}
2 \int_{t_2}^{t_1}  dt_{3} \int_{t_2}^{t_3} dt_4\,
\left[e^{-i \wb t_{34}} - e^{i \wb \tildet_{34}} \right]
\bP^\crw_{(r_{12}, t_{12})} (\bbmq, t_{34}) \; . 
 \end{eqnarray}
  \end{subequations}
\end{widetext}
  \Eq{eq:freq-to-time-result} follows by transforming the free
  Cooperons in \Eq{eq:Plongdefine} to the time domain, and
  recognizing the $(d \bmq)$ integral of the resulting expression to
  contain the object 
\begin{eqnarray}
 \nonumber 
 \bP^\crw_{(r_{12}, t_{12})}(\qb , t_{34})
& = & 
{\int (dq) \, e^{i q r_{12}} \, 
\bC^0_{q}(t_{13})\,\bC^0_{q-\qb}(t_{34})\,\bC^0_{q}(t_{42})
\over  \tCO (r_{12},t_{12})} 
\\ & = & \label{eq:IIrrw-result-explicit}
e^{ - D \qb^2 t_{34} 
\left(1 - {t_{34} \over t_{12}} \right) + i \qb r_{12}  {t_{34}
    \over t_{12}} } \, . 
\end{eqnarray}
This quantity is the Fourier transform of the probability density
$\tilde P_{(0,t)}(r',t')$ for an intermediate portion
of a  random walk to cover the distance $r'$
in the time $t'$, under the condition that the total walk is closed,
returning to the starting point $r=0$ after a total time $t$. It was
introduced in paper I [\EqsI{Uavg} and
(I.\ref{eq:rrw-result-explicit})] as central ingredient for averaging
the effective action of the influence functional derived there over
pairs of time-reversed, closed random walks.

  We are now in a position to write down an explicit
expression for the
  decay function $\Foneertt$.  Writing $ \tilde F^{\ve} = \sum_{n = 0}^\infty
  \tilde F^{\ve}_{(n)}$, we find from \Eqs{eq:Fonefirstresults} and
  \Eq{subeq:linearizedFequation}:
  \begin{eqnarray}
    \label{eq:Fonenresults}
\tilde F^{\ve}_{(n)} (r_{12}; t_1, t_2) 
&  = & {1 \over \tC^0 (r_{12}, t_{12})} \, \partial_{t_{12}}^n  
\, { 2 \over \hbar}
 \int_{t_2}^{t_1} \!\!\!  dt_{3} \! \int_{t_2}^{t_3} \!\!\!  dt_4
\qqph 
\\ \nonumber
& & \times \! 
\int (d \bbmq) \, 
\tC^0 (r_{12}, t_{12}) \, P^\crw_{(r_{12}, t_{12})} (\bbmq, t_{34}) 
\\ \nonumber
& & \times \! \int (d \bomega) \, 
 \bcL^{\deco}_{\ve (n), \bbmq} (\bomega) \! 
\left[e^{-i \wb t_{34}} \! - \! e^{i \wb \tildet_{34}} \right]
\! . 
  \end{eqnarray}
  We henceforth set $r_{12} =0$ and $t_1 = - t_2 = \toh t$, as
  required for calculating the conductivity [cf.\
  \Eq{eq:C0tfromcond}], and write $\tilde F^{\ve}_{(n)} (t) \equiv \tilde
  F^{\ve}_{(n)} (0; \toh t, - \toh t) $.  We shall discuss only the
  leading term $\tilde F^\ve_{(0)}$, since the $F_{(n>0)}$ terms give
  subleading contributions. (This is illustrated in
  Appendix~\ref{app:checkF1} for $F_{(1)}$.)  For $n = 0$, the correlator
  needed in \Eq{eq:Fonenresults} reduces [via \Eqs{eq:define:Ldeco:n}
  and (\ref{eq:define:Ldeco})] to the Pauli-principle-modified noise
  correlator of \Eq{IItheGreatReplacementa}, $\bcL^{\deco}_{\ve (n),
    \bbmq} (\bomega) = \oneoverhbar \bigl\langle \hat{V} \hat{V}
  \bigr\rangle^\pbzero_{\qb \wb}$.
After symmetrizing the range of the $t_4$ integral to
be $\int_{-t/2}^{t/2} dt_4$, and setting $t_4 \to - t_4$
in the $e^{i \wb \tildet_{34}}$ term of \Eq{eq:Fonenresults}, we
obtain
\begin{eqnarray}
\tilde F_{(0)}^\ve (t) & = & 
\oneoverhbarsq \int_{-{t \over 2} }^{{t \over 2} }  
dt_{3} \int_{-{t \over 2} }^{{t \over 2}} dt_{4}\,
\int (d\bar{q})  \int (d\bar{\omega}) \, 
e^{-i\bar{\omega}t_{34} }  \qqph 
\label{eq:exponentBetheSalpeter}
\\ \nonumber 
& & 
\times 
 \bigl\langle \hat{V} \hat{V}
  \bigr\rangle^\pbzero_{\qb \wb} 
\left[
\bP^\crw_{(0, t)}(\qb , |t_{34}|)  -
\bP^\crw_{(0, t)}(\qb , |\tildet_{34}|)  
 \right] \, .
\end{eqnarray}
This result is \emph{identical} to the function $F^{\pbzero}_\crw (t)
$ whose form was conjectured by heuristic arguments in
Section~\ref{sec:symmPauli-mod} of paper I, namely
\EqI{exponentCLpaulimodified} [with $\delta \bP $ therein given by the
``closed random walk'' version of \EqI{eq:barPPPa}]. Thus, we reach
the main conclusion of paper II: \emph{the heuristic way of
  introducing Pauli blocking into an influence functional approach} in
Paper I, Section~\ref{sec:symmPauli-mod} (and, by implication, also
the more formal analysis of Ref.~\onlinecite{vonDelft04}) \emph{is
  fully consistent with the present diagrammatic Bethe-Salpeter
  approach.}

To calculate the energy-averaged version of the Cooperon,
$\bigl \langle \tilde C^\ve (0, t) \bigr \rangle_\ve$, as needed in
\Eq{eq:finalsigmawithintmaintextpert}, we need the 
energy-average of $e^{- \tilde F_{(0)}^\ve (t) }$.
This was done in great detail in Paper I, Sec.~\ref{sec:e-averaged}
for $d=1,2,3$,
so we shall quote only the result for $d=1$ here:
\begin{eqnarray}
  \label{eq:Eaverageexponent}
  \left \langle e^{- \tilde F_{(0)}^\ve (t) } \right \rangle_\ve 
& \simeq & e^{- \langle \tilde F_{(0)}^\ve (t) \rangle_\ve } \; , 
\\
\langle  \tilde F^\ve_{(0)} (t) \rangle_\ve & = & 
\tilde F^\class_{\crw} (t) \left[ 1  + {\cal O} \bigl[ (t
  T)^{-1/2} 
\bigr] \right] \; .  \label{eq:finalresultforEFfull}
\end{eqnarray}
Here $F^\class_\crw (t) = (\sqrt {\pi}/4) (t/\tauphiAAKone)^{3/2}$,
with $\tauphiAAKone$ given by \Eq{eq:tauhphiAAK}, is the result for
the decay function $\tilde F^\class_{\crw} (t)$ obtained in paper I
[\EqI{eq:finalFrrw}] for classical white Nyquist noise.
\Eq{eq:finalresultforEFfull} states that the leading quantum
correction to the $\tilde F^\class_{\crw} (t)$ is of order $(Tt)^{-1 /
  2}\propto g^{-1 / 2}$, \ie\  small in the regime where
weak localization theory is applicable. [The numerical prefactor
of this term was evaluated explicitly in paper I, see
\Eq{eq:F1g(L_t)}].

\vspace{0.2cm} 
\subsection{Comparison with magnetoconductivity of AAG}

As a final check of our Bethe-Salpeter analysis, let us use it to
directly calculate $\delta\sigma^{\WL (1)}$, the first term in an
expansion of the weak localization conductivity in powers of the
interaction.  Inserting $\tilde C^{\ve (1)}_\deco (0; \toh t,
- \toh t)$ from \Eq{eq:C1tauexplicit} into \Eqs{eq:Ctimeintegral} and
\Eq{eq:magnetoconductancefull} to obtain $\delta\sigma^{\WL (1)}$, the
$\int d t$ integral produces a $\delta (\omega)$ that sets $\omega =
0$; after some obvious substitutions [from \Eqs{eq:Sigmabaredeco},
(\ref{eq:selfenergy-vertI-new-old}) and (\ref{IItheGreatReplacementa})]
we readily find that the resulting expression
for $\delta\sigma^{\WL (1)}$ is given precisely by
\EqI{eq:AAGresult}, which, as mentioned previously, agrees with
Eq.~(4.5) of AAG\cite{AAG}.

\subsection{Plausibility arguments for exponential Ansatz}
\label{sec:whyexpAnsatz}

To end this section, some remarks on the adequacy of our exponential
Ansatz are in order. First, if an exact solution for
\Eq{eq:eqforFfunction} for $\Fertt$ could be found, the exponential
Ansatz (\ref{eq:AnsatzCewtau}) for $\Certt$ would yield the exact
expression for the Cooperon. Of course, however, it was necessary to
make approximations in solving \Eq{eq:eqforFfunction}, and once these
have been made, one might question whether the exponential Ansatz
adequately captures the important physics.  For example, one might
consider functional forms of the type $A_d (t) e^{-\tilde F_d(t)}$, as
discussed by Golubev and Zaikin \cite{GZ3}, where the prefactor
$A_d(t)$ has a non-trivial time dependence different from $\tilde
C^0$. [With such an Ansatz, it would not be possible to determine $A_d
(t)$ and $F_d (t)$ from a first-order calculation of the Cooperon,
since it would be unclear how to separate the contributions of $A_d
(t)$ and $\tilde F_d (t)$ to $\tilde C^{1}$ in order to decide which
part has to be reexponentiated and which part should stay in the
prefactor.] Indeed, GZ have argued \cite{GZ3} that the final
expression for $\tilde F_d(t)$ after averaging over diffusive paths
contains only $\coth [\bomega/2T]$ factors and no $\tanh [(\ve \mp
\bomega)/2T]$ factors, and that the latter instead only contribute to
the prefactor $A_d (t)$, in such a way that an expansion of $A_d (t)
e^{-\tilde F_d(t)}$ to first order in the interaction propagator
correctly reproduces the combinations $\coth + \tanh$ occuring in
$\tC^{1}$. In our language, that would correspond to reexponentiating
only contributions from $\bSigma{}^{ I }$, while attributing all
contributions from $\bSigma{}^{ R }$ to the prefactor $A_d (t)$.

However, there are several strong arguments against such a procedure.
Firstly, the diagrams for $\bSigma{}^{ I }$ and $\bSigma{}^{ R }$
always occur in matching pairs, generating a series of products of the
type $\bcC^{(0)} \bigl([\bSigma^I + \bSigma^R ] \bcC^{(0)} \dots
\bcC^{(0)} \bigl([\bSigma^I + \bSigma^R ] \bcC^{(0)} $. An Ansatz
assigning $\bSigma^R$ to the prefactor only and $\bSigma_I$ to the
exponent would disrespect the structure of this series. Secondly, the
structure of the Ansatz should be sufficiently general that it holds
for any dimensionality, $d=1,2,3$. But for $d=3$, the complicated
Bethe-Salpeter analysis is not necessary and the Dyson equation
sufficient, because the self-energy contributions
$\bSigma{}^{\self}_{\ve, \bmq} (\omega) $
[\Eq{eq:effectiveCooperonselfenergy}] are infrared-convergent by
themselves, without the need for vertex corrections; in this case, the
Cooperon decay is indeed purely exponential, $e^{- \gamma^{\varphi,
    \self}_{0, 0} t}$, where the decoherence rate is given by $
\gamma^{\varphi, \self}_{0, \ve} $ of \Eq{eq:gamma0selffirstexp},
which evidently \emph{does} contain the combination $\coth (\bomega/
2T ) + \tanh (\varepsilon - \bomega )/ 2T $.  The fact that this
combination shows up in $ \tilde F_3 (t)$ implies that it should also
show up in 2 and 1 dimensions for the decay functions $\tilde F_{1,2}
(t)$.  Thirdly, in the limit $\ve \gg T$, it is general consensus that
the decoherence rate is simply given by the inelastic rate, $F^\ve_d
(t) \sim \ve^{d/2}t$, and indeed this result was recovered from our
theory in paper~I [Sec.~\ref{sec:Energy-dep-decay-function}]; but this
is possible only if $\tilde F^\ve (t)$ contains $\tanh[(\ve -\wb/2T)]$
functions, since they are the only way in which the energy dependence
enters the theory.  And finally, the fact that the combination $\coth
+ \tanh $ occurs in the effective action of the influence functional,
\ie\ in the exponent, was derived by GZ themselves \cite{GZ2} using
their influence functional approach (the fact that the $\tanh$-terms
dropped out of their final results for the decoherence rate is only
due to their neglect of recoil, as shown in \Ref{vonDelft04}).

\section{Conclusions}

In papers I and II, we have shown how the combined effects of quantum
noise and the Pauli principle can be incorporated into a calculation
of decoherence rate of interacting electrons in disordered metals. To
this end, we used both an influence functional formulation and
standard diagrammatic methods, obtaining identical results with both
methods.  The influence functional approach is perhaps more
intuitively transparent: it is formulated in the position-time domain,
where we have intuition about the behavior of diffusive trajectories,
and shows very nicely how for quantum noise the contribution to
decoherence that arises from spontaneous emission gets cancelled by
Pauli blocking at sufficiently low temperatures. Thus, we find that as
long as the condition $T \tauphi \gg 1$ holds (which characterizes the
regime of weak localization), the decoherence rate decreases without
saturation as the temperature is lowered towards zero. The fact that
Golubev and Zaikin obtain a saturation was identified to be due to
their neglect of recoil.  

However, paper I does rely on heuristic
arguments in the way Pauli blocking is introduced.  Corroborating the
correctness of these arguments was the purpose of paper II; and
indeed, setting up a Bethe-Salpeter and solving it by an exponential
Ansatz, we recovered the decay function found in paper I. Moreover, we
identified several correction terms that do not arise in the influence
functional approach ($F^{(1)}$, $\tilde{C}^{(1) \ve}_{\self, Z}$) and
showed them to be negligble.

Apart from clarifying the fundamentally important interplay between
spontaneous emission and Pauli blocking, our calculation of the
decoherence rate has the merit of being free from any infrared or
ultraviolet divergencies: in the leading terms that govern the
decoherence rate, all necessary cutoffs arise naturally from within
our formalism, and do not need to be inserted by hand (whereas AAK did
need to insert an UV cutoff by hand for $d=2,3$).  This has enabled us
to obtain a number of new results.  Firstly, our more accurate
treatment of the regime of large frequency transfers ($\bomega \simeq
T$) has allowed us to calculate explicitly the leading quantum
corrections to the results of AKK for $\tauphi$
[\EqsI{subeq:explicitshifts}], finding them to be small in $1/T
\tauphi$. Secondly, by explicitely keeping track of the energy
dependence of the propagation energy $\ve$ of the diffusing electrons,
we were also able to discuss in detail the energy-dependence of the
decoherence rate, also for energies higher than the temperature
\EqsI{subeq:Fcrossover}.

\begin{acknowledgments} 
  We thank I. Aleiner, B. Altshuler, M. Vavilov, I. Gornyi, and in
  equal measure D. Golubev and A. Zaikin, for numerous patient and
  constructive discussions.  Moreover, we acknowledge illuminating
  discussions with J. Imry, P. Kopietz, J. Kroha, A.  Mirlin, G.
  Montambaux, H.  Monien, A. Rosch, I.  Smolyarenko, G. Sch\"on,
  W\"olfle and A.  Zawadowski.  Finally, we acknowledge the
  hospitality of the centers of theoretical physics in Trieste, Santa
  Barbara, Aspen, Dresden and Cambridge, and of Cornell University,
  where some of this work was performed.  F.M.\ acknowledges support
  by a DFG scholarship (MA 2611/1-1), V.A.\ and R.S.\ support by NSF
  grant No. DMR-0242120.

\end{acknowledgments}


\appendix

\begin{widetext}

\section{Diagrammatic derivation of 
Bethe-Salpeter equation}
\label{app:BS}

\begin{figure}[htbp]
\begin{centering}
\includegraphics[width=0.9\linewidth]{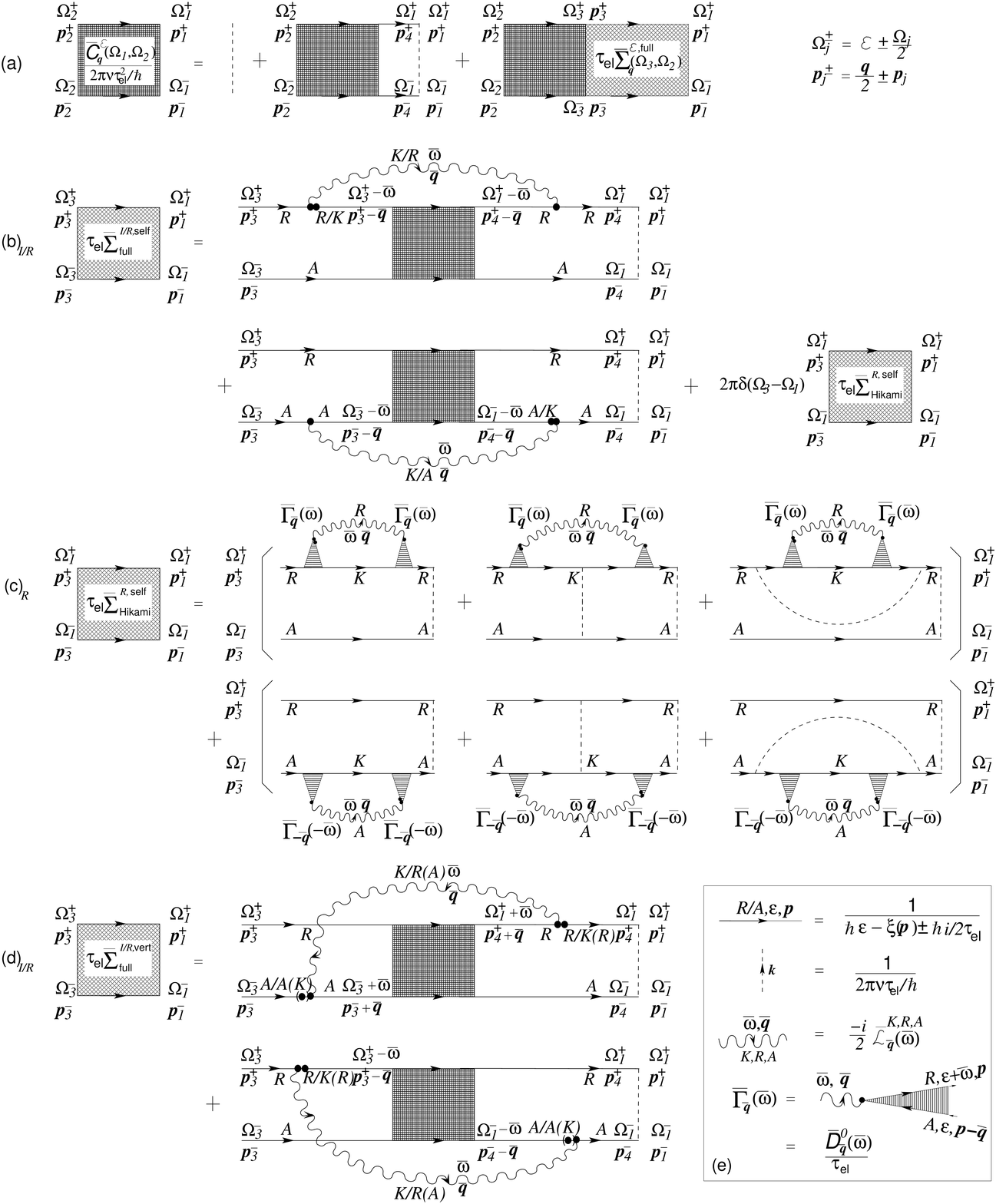}
\end{centering}
\caption{(a) Diagrammatic depiction of the Bethe-Salpeter equation
  \Eq{eq:Bethe-Salpeter-full} for the Cooperon $ \bar {\cal
    C}_\bmq^{{\cal E}} (\Omega_1, \Omega_2)$.  (b), (c) and (d): The
  contributions to the full self-energy $\bSigma{}^{\cal E}_{\bmq,
    \full} (\Omega_1, \Omega_3)$ of
  \Eq{eq:Bethe-Salpeter-full-selfenergy}, namely (b):
  $\bSigma{}^{{\cal E}, I/R, \self}_{\bmq, \bbmq, \full} (\Omega_1,
  \Omega_3, \bomega) $; (c) $\bSigma{}^{{\cal E}, R }_{\bmq, \bbmq,
    \Hikami} (\Omega_1, \bomega) $ (which contributes to
  $\bSigma{}^{R, \self}_{\full} $ only); and (d) $\bSigma{}^{{\cal E},
    I/R, \vertex}_{\bmq, \bbmq, \full} (\Omega_1, \Omega_3, \bomega)
  $.  The superscripts $I/R$ on $\bSigma{}$ indicate which type of
  interaction propagator occurs: $\bSigma{}^I$ (generated by $S_I$ in
  GZ's approach\cite{GZ2,vonDelft04}) contains $\barLL^{K}$, together
  with $\bcG{}^R$ and $\bcG{}^A$ on the upper and lower contours,
  respectively; $\overline \Sigma^R$ (generated in GZ's approach by
  $S_R$) contains a sum of two types of terms, featuring the
  combinations $\barLL^{R}$ together with $\bcG{}^K$ and $\bcG{}^A$ on the
  upper and lower contours, or $\barLL^A$ together with $\bcG{}^R$ and
  $\bcG{}^K$ on the upper and lower contours, respectively.  The diagrams
  in (d) for $\bSigma{}^{I/R, \vertex}_{\full} $ occur both without
  and with bracketed labels for the interaction and electron
  propagators, \eg\ $R$ or $(A)$ on ${\cal L}$, to distinguish
  contributions in which the electron Keldysh Green's function occurs
  on the upper or lower electron line, respectively.  The vertex to
  which the electron Keldysh Green's function is attached is always
  indicated by a double dot (adopting a convention used in
  \cite{vonDelft04}). (e) Diagrammatic conventions; the dressed
  interaction vertex is denoted by $\bar \Gamma_\qb (\wb) =
  [\bcD^0_{\bbmq} (\bomega) ]/\tauel$, where $\bcD^0$ is the bare
  diffuson propagator.}
\label{fig:bethesalpeter}
\end{figure}
\end{widetext}

This appendix diagrammatically specifies the Bethe-Salpeter equation
governing the Cooperon, and gives explicit expressions
for the full and bare Cooperon self-energies occuring
$\rule[-4mm]{0mm}{0mm}$therein. 

\subsection{Full Bethe-Salpeter equation}
\label{app:BSdiagrams}

 The diagrams specifying the Bethe-Salpeter equation for the
full Cooperon, first written down in Ref.~\onlinecite{AAV}, 
are shown in Fig.~\ref{fig:bethesalpeter}. 
The feature that distinguishes
``weak-localization'' from ``interaction'' corrections to the
conductivity is that for the former each current vertex is attached to
a retarded and an advanced electron propagator, $G^R j G^A$, whereas
for the latter, one or both current vertices are connected to two
retarded or two advanced electron lines, $G^{R} j G^{R}$ or $G^{A} j
G^{A}$. Hence, in setting up the Bethe-Salpeter equation, only those
types of diagrams have been included for which the upper (or lower)
lines entering and leaving the Cooperon are both retarded (or
advanced) electron propagators. By adopting a pure ladder structure in
Fig.~\ref{fig:bethesalpeter}(a), diagrams containing overlapping
interaction lines have been dropped, but these are known 
\cite{GrilliSorella88}
to be smaller than those included by at least a factor of $(\kF
\lel)^{-1} \ll 1$ (where $\lel$ is the mean free path).

Fig.~\ref{fig:bethesalpeter}(a) translates into the following equation:
\begin{widetext}
\begin{eqnarray}
\label{eq:Bethe-Salpeter-full}
{1 \over \tauel} \bcCEq (\Omega_1, \Omega_2) & = & 
2 \pi \delta (\Omega_{12}) +
\left[ {1 \over \tauel} - (E_\bmq - i \Omega_1) \right]
\bcCEq (\Omega_1, \Omega_2) 
+ \int (d \Omega_3) \, 
\bSigma{}^{\cal E}_{\bmq, \full} (\Omega_1, \Omega_3) \, 
\bcCEq (\Omega_3, \Omega_2) \; . 
\end{eqnarray}
This equation is equivalent to \Eq{eq:Bethe-Salpeter-full-compact}
of the main text. 

The ``Cooperon self-energy'' $\bSigma{}^{\cal E}_{\bmq, \full}$
occuring herein is defined by
\begin{subequations}
  \label{subeq:newselfenergies}
\begin{eqnarray}
\label{eq:Bethe-Salpeter-full-selfenergy}
\bSigma{}^{\cal E}_{\bmq, \full} (\Omega_1, \Omega_3)  & = &
{ 1 \over \hbar} \int (d \bbmq) \int (d \bomega)  
\left[
\bSigma{}^{{\cal E}, I +R , \self}_{\bmq, \bbmq, \full} 
(\Omega_1, \Omega_3,  \bomega) +
\bSigma{}^{{\cal E}, I + R, \vertex}_{\bmq, \bbmq, \full} 
(\Omega_1, \Omega_3,  \bomega) 
\right] \, , \qqph 
\end{eqnarray}
where the subscript $\bSigma{}^{I+R}$ indicates a sum $\bSigma{}^I +
\bSigma{}^R$, and the self-energy and vertex contributions to $\bSigma{}$
are given diagrammatically by Figs.~\ref{fig:bethesalpeter}(b) to
\ref{fig:bethesalpeter}(d). These lead to the following expressions
(here $\Omega_j^\pm \equiv {\cal E} \pm \toh \Omega_j$):
 \begin{eqnarray}
  \label{eq:selfenergy-selfI-new}
\bSigma{}^{{\cal E}, I , \self}_{\bmq, \bbmq, \full} 
(\Omega_1, \Omega_3,  \bomega) 
 & \equiv &  
\phantom{-} \toh i \bcL^K_\bbmq (\bomega) 
 \left[ 
\bcC^{{\cal E} - {1 \over 2} \bomega}_{\bmq - \bbmq}
( \Omega_1 - \bomega,  \Omega_3 - \bomega) +
\bcC^{{\cal E} - {1 \over 2} \bomega}_{\bmq - \bbmq}
( \Omega_1 + \bomega,  \Omega_3 + \bomega) \right] \; , 
\rule[-5mm]{0mm}{0mm}
\\
 \label{eq:selfenergy-selfR-new}
\bSigma{}^{{\cal E}, R , \self}_{\bmq, \bbmq, \full} 
(\Omega_1, \Omega_3,  \bomega) 
& \equiv &
\phantom{-} 
\tanh [(
\Omega_3^+ - \bomega)/2T] \,
\toh i \bcL^R_\bbmq (\bomega) \, 
\bcC^{{\cal E} - {1 \over 2} \bomega}_{\bmq - \bbmq}
( \Omega_1 - \bomega,  \Omega_3 - \bomega) \\
\nonumber
& & -
\tanh [ (
\Omega_1^- - \bomega)/2T] \,
\toh i \bcL^A_\bbmq (\bomega) \, 
\bcC^{{\cal E} - {1 \over 2} \bomega}_{\bmq - \bbmq}
( \Omega_1 + \bomega,  \Omega_3 + \bomega)  
\\
\nonumber
& & 
 + \;  2 \pi \delta (\Omega_1 - \Omega_3) \, 
\bSigma{}^{{\cal E}, R }_{\bmq, \bbmq, \Hikami} 
(\Omega_1,   \bomega) \; , 
\rule[-5mm]{0mm}{0mm}
\\
\label{eq:Hikamiself}
\bSigma{}^{{\cal E}, R }_{\bmq, \bbmq, \Hikami} 
(\Omega_1,   \bomega) 
& \equiv & 
\phantom{-} 
\tanh [ (
\Omega_1^+ - \bomega)/2T] \,
\toh i \bcL^R_\bbmq (\bomega) \, 
\bigl[ \bcD^0_\bbmq(\bomega) \bigr]^2
\left( \bigl[ \bcC^0_{\bmq} ( \Omega_1 ) \bigr]^{-1}
+ \bigl[ \bcD^0_{\bbmq} (\bomega) \bigr]^{-1} \right)
 \qqph 
\\ \nonumber
& &  
  - \tanh [ (
\Omega_1^- - \bomega)/2T] \,
\toh i \bcL^A_\bbmq (\bomega) \, 
\bigl[ \bcD^0_{-\bbmq} (- \bomega) \bigr]^2
\left( \bigl[ \bcC^0_{\bmq} ( \Omega_1 ) \bigr]^{-1}
+ \bigl[\bcD^0_{\bbmq} (- \bomega) \bigr]^{-1}
\right) , 
\rule[-5mm]{0mm}{0mm}
\\
  \label{eq:selfenergy-vertI-new}
\bSigma{}^{{\cal E}, I , \vertex}_{\bmq, \bbmq, \full} 
(\Omega_1, \Omega_3,  \bomega) 
& \equiv &
- \toh i \bcL^K_\bbmq (\bomega) 
 \left[ 
\bcC^{{\cal E} + {1 \over 2} \bomega}_{\bmq + \bbmq}
( \Omega_1 + \bomega,  \Omega_3 - \bomega) +
\bcC^{{\cal E} - {1 \over 2} \bomega}_{\bmq - \bbmq}
( \Omega_1 + \bomega,  \Omega_3 - \bomega) \right]  \; ,  
\rule[-5mm]{0mm}{0mm}
\\
 \label{eq:selfenergy-vertR-new} 
\bSigma{}^{{\cal E}, R , \vertex}_{\bmq, \bbmq, \full} 
(\Omega_1, \Omega_3,  \bomega) 
& \equiv &
\phantom{-} 
\left[
\tanh [ \Omega_1^+ 
 /2T] \,
\bcL^R_\bbmq (\bomega)
-
\tanh [ \Omega_3^- 
/2T]  \, \bcL^A_\bbmq (\bomega)
\right] \! 
(-\toh i ) \, \bcC^{{\cal E} + {1 \over 2} \bomega}_{\bmq + \bbmq}
( \Omega_1 + \bomega,  \Omega_3 - \bomega) 
\\
 \nonumber
&& 
+ \left[
\tanh [ (\Omega_3^+ 
- \bomega ) /2T] \,
 \bcL^R_\bbmq (\bomega)
-
\tanh [ (\Omega_1^- 
- \bomega ) /2T] 
\,  \bcL^A_\bbmq (\bomega) \right] \! 
(- \toh i ) \, 
\bcC^{{\cal E} - {1 \over 2} \bomega}_{\bmq - \bbmq}
( \Omega_1 + \bomega,  \Omega_3 - \bomega) 
\, . 
\end{eqnarray}
\end{subequations}
\end{widetext}
The terms in \Eq{eq:Hikamiself}, with their characteristic dependence
on $\bigl( \bcD^0 \bigr)^2 \bigl[ \bigl(\bcC^0 \bigr)^{-1} +
\bigl(\bcD^0 \bigr)^{-1} \bigr]$, stem from the Hikami-box
contributions of Fig.~\ref{fig:bethesalpeter}(c). The ingredients
entering the above equations are given by
\begin{subequations}
\label{subeq:ingredients}
\begin{eqnarray}
 \bLKbbqw & = & 2 i \coth ( \bomega  /2T) \, \textrm{Im}
\bigl[ \bcL^R_\bbmq (\bomega) \bigr] \; , 
\\ 
  \label{eq:recallLCD}
  \bLRbbqw & = & [\bLAbbqw]^\ast 
\simeq
- {E^0_\bbmq - i \bomega  \over  2\nu E_\bbmq^0}
\\
\bcC^0_\bbmq (\bomega) & =&  {1 \over E_\bbmq - i \bomega}
\; ,
\qquad
\bcD^0_\bbmq (\bomega) = {1 \over E_\bbmq^0 - i \bomega }
\; , \qqph 
\\
\label{eq:defineEq}
E^0_\bbmq  & \equiv & \Dd \bbmq^2 \; , \qquad 
E_\bbmq \equiv \Dd \bbmq^2 + \gamma_H \; . 
\end{eqnarray}
\end{subequations}
In \Eq{eq:recallLCD} we have taken 
the usual ``unitary limit'',
which is relevant in the limit of small frequencies
and momenta; the more general expression is\cite{AltshulerLarkin82}
\begin{eqnarray}
  \label{eq:generalinteraction}
  \bLRbbqw & = & 
  - \; {E^0_\bbmq - i \bomega\over  
        2 \nud E^0_\bbmq + ( E^0_\bbmq - i \bomega ) / V^{(d)}_\bbmq }
\; , 
\end{eqnarray}
where $V^{(d)}_\bbmq = a^{3-d} \int d^d r \, e^{-i \bbmq r} (e^2/r) $ 
is the Fourier transform of 
the Coulomb potential in $d$ effective 
dimensions:
$
V^{(3)}_\bbmq  =  {e^2 \, 4 \pi / \bbmq^2}
$, 
$V^{(2)}_\bbmq  =  {a\, e^2 \,  2 \pi / |\bbmq| }$, 
and 
$V^{(1)}_\bbmq  = a^2 e^2  \ln (\bbmq^2 a^2)$\rule[-5mm]{0mm}{0mm}. 

\subsection{Bare Self-Energies}
\label{sec:barSigma}

Since the self-energies $\bSigma{}_\full$ of
\Eqs{subeq:newselfenergies} are proportional
to the Cooperon $\bcC$, the Bethe-Salpeter equation
(\ref{eq:Bethe-Salpeter-full-compact}) is non-linear in $\bcC$. 
Solving it in its full glory thus seems hardly feasible. Thus,
we shall ``linearize'' it 
%
by making the replacement 
$\bcC^{\cal E }_{\bmq } (\Omega_1 , \Omega_3) \to 
2 \pi \delta (\Omega_{13}) \, \bcC^0_\bmq (\Omega_1) $ for
every occurence of the full Cooperon in the self-energy terms. The
resulting bare self-energies can be written in the form 
\begin{widetext}
\begin{subequations}
\label{subeq:bareselfenergies}
\begin{eqnarray}
\label{eq:Bethe-Salpeter-new-selfenergy}
\bSigma{}^{\cal E}_{\bmq, \bare} 
(\Omega_1, \Omega_4) & = &
{ 1 \over \hbar} \int (d \bbmq) 
\int (d \bomega) \, 2 \pi
\left[
  \delta (\Omega_{14})  \,
  \bSigma{}^{{\cal E} + {1 \over 2} \Omega_4,
    \self}_{\bmq, \bbmq, \bare} (\widetilde \Omega_{14}, \bomega) 
  \, + \, 
  \delta (\Omega_{14} + 2 \bomega)  \, 
  \bSigma{}^{{\cal E} + {1 \over 2} \Omega_4,
    \vertex}_{\bmq, \bbmq, \bare} (\widetilde \Omega_{14} , \bomega ) 
\right]  , \qqph 
\end{eqnarray}
where $\Omega_{14} \equiv \Omega_1 - \Omega_4$ and $\widetilde
\Omega_{14} \equiv \toh (\Omega_1 + \Omega_4)$, with the following
ingredients:
  \label{subeq:newselfenergies-old}
 \begin{eqnarray}
  \label{eq:selfenergy-selfI-new-old}
\bSigma{}^{{\cal E}, \self}_{\bmq, \bbmq, \bare} (\omega, \bomega)  & = &
\left[\bSigma{}^{{\cal E}, \self, \deco}_{\bmq, \bbmq, \bare} + \bSigma{}^{{\cal E},
    \self,
    Z}_{\bmq, \bbmq, \bare}\right] (\omega, \bomega) \; , 
\rule[-5mm]{0mm}{0mm}
\\
\label{eq:Sigmabaredeco}
\bSigma{}^{{\cal E} ,\self, \deco }_{\bmq, \bbmq, \bare} 
(\omega, \bomega)
 & \equiv &   -  \left[ \bcC^0_{\bmq - \bbmq}( \omega - \bomega) +
\bcC^0_{\bmq - \bbmq} (\omega + \bomega)  \right] 
\bcL^{\deco}_{{\cal E} \omega, \bbmq} (\bomega) \; , 
\rule[-5mm]{0mm}{0mm}
\\
\label{eq:SigmabareZ}
 \bSigma{}^{{\cal E} , \self, Z}_{\bmq, \bbmq, \bare} (\omega, \bomega) 
& \equiv &
\phantom{+} 
\tanh [ ({\cal E} - \bomega)/2T] \,
\toh i \bcL^R_\bbmq (\bomega) 
\left[\left[ \bcD^0_\bbmq(\bomega)  \right]^2  
\Bigl(
\bigl[ \bcC^0_{\bmq} ( \omega ) \bigr]^{-1}
+ \bigl[ \bcD^0_{\bbmq} (\bomega) \bigr]^{-1}
\Bigr) -  \bcC^0_{\bmq - \bbmq} (\omega + \bomega)  
\right]
\\ \nonumber
& & -
\tanh [ ({\cal E} - \bomega - \omega)/2T] \,
\toh i \bcL^A_\bbmq (\bomega) 
\left[\left[ \bcD^0_\bbmq(-\bomega)  \right]^2  
\Bigl(
\bigl[ \bcC^0_{\bmq} ( \omega ) \bigr]^{-1}
+ \bigl[ \bcD^0_{\bbmq} (-\bomega) \bigr]^{-1}
\Bigr) -  \bcC^0_{\bmq - \bbmq} (\omega - \bomega)  
\right] \; , \qquad 
\rule[-5mm]{0mm}{0mm}
\\
  \label{eq:selfenergy-vertI-new-old}
\bSigma{}_{\bmq, \bbmq, \bare}^{{\cal E}, \vertex} 
(\omega, \bomega) & \equiv &
2  \bcC^0_{\bmq - \bbmq}( \omega ) \, 
\bcL^{\deco}_{{\cal E} \omega, \bbmq} (\bomega)  \; , 
\rule[-5mm]{0mm}{0mm}
\\
\label{eq:define:Ldeco}
\bcL^{\deco}_{{\cal E} \omega, \bbmq} (\bomega) & \equiv &
\coth [ \bomega/2T] \, \textrm{Im}
\bigl[ \bcL^R_\bbmq (\bomega) \bigr] 
- \tanh [ ({\cal E} - \bomega)/2T] \,
\toh i \bcL^R_\bbmq (\bomega) +
\tanh [ ({\cal E} - \bomega - \omega)/2T] \,
\toh i \bcL^A_\bbmq (\bomega) \; . \qquad 
\end{eqnarray}
\end{subequations}
\end{widetext}
We have split the self-energy contribution [stemming from
\Eqs{eq:selfenergy-selfI-new} plus (\ref{eq:selfenergy-selfR-new})]
into two terms, $\bSigma{}^{\self, \deco}_\bare + \bSigma{}^{\self,
  Z}_\bare $, chosen such that the Hikami-box contributions are fully
contained in $\bSigma{}^{\self, Z}_\bare $, and that both
$\Sigma^{\self, \deco}_\bare$ and $\bSigma{}^\vertex_\bare$ are
\emph{proportional to the same combination of propagators,}
$\bcL^{\deco} $ [\Eq{eq:define:Ldeco}], a feature that considerably
simplifies the analysis in the main text.  To achieve this, the terms
in \Eq{eq:SigmabareZ} that are proportional to $\tanh \cdot \,
\bcC^0_{\bmq - \bbmq} (\omega \pm \bomega) $ were added in
\Eq{eq:Sigmabaredeco} and subtracted in (\ref{eq:SigmabareZ}),
respectively.  This addition-subtraction trick amounts to
''replacing'' the Hikami-box contribution to $\Sigma^{\self,
  \deco}_\bare$ by ``replacement terms'' [those added to
\Eq{eq:Sigmabaredeco}] that (i) have a simpler, more convenient
structure (since proportional to $\bcC^0$ instead of $\bigl( \bcD^0
\bigr)^2 \bigl[ \bigl(\bcC^0 \bigr)^{-1} + \bigl(\bcD^0 \bigr)^{-1}
\bigr]$), but (ii) nevertheless have the same leading infrared and
ultraviolet behavior, in the sense that the difference between the
Hikami-box and the replacement terms, namely $\bSigma{}^{\self, Z} $,
generates only subleading contributions to the long-time behavior of
the Cooperon (as explained below). The leading contribution comes from
$\bSigma{}^{\self, \deco} $ and $\bSigma{}^{\vertex} $,
because both are proportional to the effective propagator
$\bcL^{\deco}_{{\cal E} \omega, \bbmq} (\bomega) $, whose $\coth
[ \bomega/2T]$ term at small frequencies $\bomega$ makes the
dominant contribution [in contrast, $\bSigma{}^{\self, Z} $ lacks such
a term]. 

Although the dominant contribution comes from low frequencies,
$\bSigma{}^{\self, \deco} +\bSigma{}^{\vertex} $ contain no infrared
divergence, since for $\bomega \to 0$, their contributions cancel each
other, as is clear directly from \Eqs{eq:Sigmabaredeco} and
(\ref{eq:selfenergy-vertI-new-old}) (or by inserting them into
\Eqs{eq:Cwbomega1} and (\ref{eq:Cwbomega2}) below, or
\Eq{eq:C1tauexplicit} of the main text).  Moroever, the dominant
contribution from $\bSigma{}^{\self, \deco} +\bSigma{}^{\vertex} $
also contains no ultraviolet divergencies, since the effective
propagator $\bcL^{\deco}_{{\cal E} \omega, \bbmq} (\bomega) $
evidently vanishes exponentially in the limit $\bomega \gg {\cal E},
T$, hence the $\int (d \bomega) $ integrals over both
$\bSigma{}^{{\cal E} ,\self, \deco }_{\bmq, \bbmq, \bare} (\omega,
\bomega) $ and $\bSigma{}^{{\cal E} ,\vertex}_{\bmq, \bbmq, \bare}
(\omega, \bomega) $ are separately free from ultraviolet divergencies. 

Concerning the contribution from $\bSigma{}^{\self, Z} $, there are
several ways to convince oneself that its contribution to the Cooperon
decay function $F(t)$ is subleading.  Firstly, an explicit
calculation, performed in Appendix~\ref{sec:SigmaZsubleading}, shows
that its contribution to $\tilde C^{(1)\ve}$, the first-order
expansion of the Cooperon in powers of the interaction propagator,
depends much more weakly on propagation time $t$ than that from
$\bSigma{}^{\self, \deco} + \bSigma{}^{\vertex} $; for example, for
quasi 1-dimensions it scales with $t^{-1/2} $ [cf.\ \Eq{eq:C1fromSigmaZ}],
compared to the $ t^{1}$ of the leading terms 
$\tC^0 (0,t) \tilde F^\ve (t) $.

An alternative, and more simple, argument goes as follows: according
to the short-cut ``self-energy-diagrams-only'' approach of
Section~\ref{sec:standardDyson}, the decoherence rate is given by
$\gamma_{\ve, 0}^{\varphi,\self} $ of \Eq{eq:definegammaphi} [see also
(\ref{eq:defineFeself})], for which we have to set $E_\bmq = i \omega$
and take $q \to 0$. Now, in this limit $\bSigma{}^{\self,Z}_{\ve,
  \bmq} (\omega)$ \emph{vanishes identically}, regardless of the form
of the interaction propagator $ \bLRbbqw$, and hence does not
contribute to $\gamma_{\ve, 0}^{\varphi,\self} $ at all.  Actually, an
even stronger statement can be made if the general form
(\ref{eq:generalinteraction}) for the interaction propagator is
specialized to the the so-called ``unitary limit'' of
\Eq{eq:recallLCD} (as is usually done anyway).  In that case, both the
Hikami-box terms and the ``replacement terms'' \emph{separately}
vanish in the limit $E_\bmq = i \omega$ and $q \to 0$.  Thus, for the
``unitary limit'' of the interaction propagator, Hikami-box
contributions actually do not contribute to the decoherence rate at
all\cite{vonDelft04}. It is for this (somewhat fortuitous) reason that
the influence functional theory of decoherence developed in
Ref.~\onlinecite{vonDelft04} was able to correctly obtain the Cooperon
decay function, despite of the fact that it did not include any
Hikami-box contributions. 

\subsection{Expansion to 2nd order in the interaction}
\label{sec:2ndorderexpansion}

The object needed on the right-hand side of \Eq{eq:Cetildewtaunew} 
for $\tilde C^\ve (r_{12},t_1,
t_2)$ is $ \bcC^{\ve - {1 \over 2} \Omega_2}_{\bmq} (\Omega_1 ,
\Omega_2) $, which is determined by the Bethe-Salpeter equation
\Eq{eq:Bethe-Salpeter-full-compact}, with ${\cal E} = \ve - \toh
\Omega_2$ (with $\wz=0$ here). 
It is instructive to consider the first few terms that are obtained
upon iterating this equation, while using the bare self-energies of
\Eq{eq:Bethe-Salpeter-new-selfenergy}. (Of course, as soon as we go
beyond first order, we should not use the bare self-energy, but the
full one, which should be calculated iteratively order by order, too;
we shall refrain from doing so here, since our intention is merely to
illustrate the general structure of the terms arising in second and
higher orders, not to evaluate them explicitly.) 
$\bcC^{\ve - {1 \over 2} \Omega_2}_{\bmq} (\Omega_1 , \Omega_2) =
\bcC^{(0)} + \bcC^{(1)} + \bcC^{(2)}$, and evaluating the result for
$\Omega_{1,2} = \omega \mp \tomega$, we obtain:
\begin{widetext}
\begin{subequations}
\label{subeq:iterateBS}
\begin{eqnarray}
\label{eq:Cwbomega0}
  \bcC^{(0)} & = & \bcC^0_\bmq (\omega) \, 
\delta (2 \tomega) \; , 
\\
\label{eq:Cwbomega1}
\bcC^{(1)} & = & 
  { 1 \over \hbar} \int (d \bbmq_1)  (d \bomega_1)  
\Biggl\{  \delta (  2 \tomega) \, 
 \bcC^0_\bmq (\omega ) \,
\bSigma{}^{\ve , \self}_{\bmq, \bbmq_1, \bare} 
(\omega , \bomega_1) \, 
  \bcC^0_\bmq (\omega ) \, 
\Biggr. 
\\ \nonumber
& & 
\qquad 
  \Biggl. 
+ \, \delta (2 \tomega - 2 \bomega_1 ) \, 
 \bcC^0_\bmq (\omega - \bomega_1) \,
\bSigma{}^{{\ve},  \vertex}_{\bmq, \bbmq_1, \bare} 
(\omega , \bomega_1) \,  \bcC^0_\bmq (\omega + \bomega_1) 
\Biggr\} \; , 
\\
\label{eq:Cwbomega2}
\bcC^{(2)} &  = &  
 { 1 \over \hbar^2} \int (d \bbmq_1)  (d \bomega_1)  (d \bbmq_2)  (d \bomega_2)  
\Biggl\{ 
\delta (2 \tomega ) \, 
 \bcC^0_\bmq (\omega ) \,
\bSigma{}^{{\ve }, \self}_{\bmq, \bbmq_1, \bare} 
(\omega , \bomega_1) \, 
\bcC^0_\bmq (\omega ) \,
\bSigma{}^{\ve , \self}_{\bmq, \bbmq_2, \bare} 
(\omega , \bomega_2) \, 
  \bcC^0_\bmq (\omega ) 
\Biggr. \qqph
\\ \nonumber
& & 
+ \, \delta (2 \tomega -   2 \bomega_1) \, 
 \bcC^0_\bmq (\omega - \bomega_1) \,
\bSigma{}^{\ve , \vertex}_{\bmq, \bbmq_1, \bare} 
(\omega , \bomega_1) \, 
\bcC^0_\bmq (\omega + \bomega_1) \,
\bSigma{}^{\ve , \self}_{\bmq, \bbmq_2, \bare} 
(\omega  + \bomega_1, \bomega_2) \, 
  \bcC^0_\bmq (\omega + \bomega_1) \, 
\rule{0mm}{5mm} 
\\ \nonumber
& & 
+ \, \delta (2 \tomega -   2 \bomega_2) \, 
 \bcC^0_\bmq (\omega - \bomega_2) \,
\bSigma{}^{\ve -  \bomega_2 , \self}_{\bmq, \bbmq_1, \bare} 
(\omega - \bomega_2, \bomega_1) \, 
\bcC^0_\bmq (\omega - \bomega_2) \,
\bSigma{}^{\ve, \vertex}_{\bmq, \bbmq_2, \bare} 
(\omega  , \bomega_2) \, 
  \bcC^0_\bmq (\omega + \bomega_2) \, 
\rule{0mm}{7mm}  \qqph 
\\ \nonumber
& & 
+ \, \delta (2 \tomega - 2 \bomega_1 - 2 \bomega_2) \, 
 \bcC^0_\bmq (\omega - \bomega_1 - \bomega_2) \,
\bSigma{}^{\ve - \bomega_2 , \vertex}_{\bmq, \bbmq_1, \bare} 
(\omega  - \bomega_2, \bomega_1) \, 
\bcC^0_\bmq (\omega + \bomega_1 - \bomega_2) \,
\rule{0mm}{7mm}
\\ \nonumber
& & \Biggl. 
\qquad \times  
\bSigma{}^{\ve , \vertex}_{\bmq, \bbmq_2, \bare} 
(\omega + \bomega_1 , \bomega_2) \, 
  \bcC^0_\bmq (\omega + \bomega_1 + \bomega_2) \, \Biggr\}
\end{eqnarray}
\end{subequations}
\end{widetext}
These expressions are useful for illustrating two important general
points.  Firstly, the expansion (\ref{subeq:iterateBS}) allows us to
confirm explicitly a fact well-known to practitioners of diagrammatic
perturbation theory, namely that the calculation of the Cooperon is
\emph{free of ultraviolet divergencies}.  This fact was implicitly
challenged by GZ, whose conclusion of a finite decoherence rate at
zero temperature stems from the occurence of an ultraviolet divergence
in their expression for the decoherence rate.  (GZ's expression
for $\gamma^\GZ$ for $d=1$ [Eq.~(76) for \Ref{GZ2}] has the form of 
our \Eq{eq:gamma0selffirstexp}, but without the $\tanh$-term, whence
they introduced an upper cut $\bomega_\textrm{max} \simeq 1/\tauel$ in
the frequency integral there; the relation of their work to
ours is discussed in more detail in Paper I, \Sec{sec:GZcomparison}.)
However, it is straightforward to check [using
\Eqs{subeq:newselfenergies-old}] that the perturbative expansion
(\ref{subeq:iterateBS}) generates no ultraviolet divergencies when
used to calculate that version of the Cooperon governing the
conductivity, namely $\tilde {\cal C}_\cond^{\ve, \wz}$ of
\Eq{eq:defineCcond}: the reason is simply that both $\bSigma{}^{\self,
  \deco}$ and $\bSigma{}^\vertex$ are proportional to the propagator
$\bcL^{\deco}_{{\cal E} \omega, \bbmq} (\bomega) $, which serves as
ultraviolet cutoff at $\bomega \simeq T$.  
[The contribution from $\bSigma{}^{\self, Z}$ is subdominant, as
mentioned above; in fact, $\bSigma{}^{{\cal E}, \self, Z}_{\bmq,
  \bbmq, \bare} (\omega, \bomega) \to 0$ in the limit of large
$\bomega$, and its leading nonzero contribution turns out to be
UV-convergent if the general expression (\ref{eq:generalinteraction})
is used for the interaction propagator, instead of its small-frequency
approximation (\ref{eq:recallLCD}).]

Secondly, we note that \emph{the frequency arguments $\bomega_i$ get
  more and more ``entangled''} from order to order in perturbation
theory, \ie, they occur in increasingly complicated combinations as
arguments of $\bSigma{}^{\self/ \vertex}$, because the vertex
diagrams cause a proliferation of frequency transfers between the
upper and lower Cooperon lines.  In $n$-th order, the generic
  structure will be
  \begin{eqnarray}
    \label{eq:Cngeneric}
     \bcC^{(n)} \sim \prod_{j=1}^n \int (d \wb_j) 
\bcC^{(0)} (\cdot) 
\bSigma ( \cdot, \wb_1) \dots
\bSigma ( \cdot, \wb_1) 
\bcC^{(0)} (\cdot) , \qph
  \end{eqnarray}
  where $(\cdot)$ stands for combinations of frequency arguments that
  can contain any number of $\wb_j$'s.  Due to this entanglement, a
  direct solution of the Bethe-Salpeter equation
  (\ref{eq:Bethe-Salpeter-full-compact}) is intractable, and further
  approximations are needed, that somehow ``factorize'' the entangled
  frequency integrals and thereby truncate the proliferation of
  frequency transfers.
  
  A natural truncation scheme would be to retain the frequency
  transfer $\bomega_j$ generated by a given vertex line only in the
  corresponding vertex function $\bSigma^{\ve, \vertex}_{\bmq,
    \bbmq_j} (\omega, \bomega_j)$, and to neglect it everywhere else
  in the diagram. As a result, it would again become possible to
  associate a definite frequency label with the upper and lower
  electron lines of the Cooperon (say $\ve $ and $\ve - \omega$).  In
  fact, such an approximation was in effect adopted in the integral
  functional approach of Ref.~\onlinecite{vonDelft04} and paper I
  (which both implicitly also took the ``long-time limit'' $\omega=0$,
  for reasons explained in the last paragraph of
  \Sec{sec:Simplified-influence-action} of paper I).  Such a procedure
  can be justified as follows: The only reason for incorporating the
  (frequency-proliferating) vertex terms in the first place, is to
  cure the infrared divergencies arising from the self-energy terms,
  which are thereby cut off at frequencies $\wb \simeq 1/t$ [this is
  perhaps seen most clearly from the $\bigl[ 1-\sin(\wb t)/(\wb t)
  \bigr]$ factor in \EqI{eq:simpleFdurw}]. For larger frequencies $\wb
  \gtrsim 1/t$, the contribution of vertex diagrams is always
  subleading compared to that of the matching self-energy diagrams,
  and hence can be neglected without affecting the leading behavior of
  the decay function $\tilde F_d(t)$.  Thus, it suffices to treat the
  $\bomega_j$-dependence associated with the frequency transfer
  between the forward and backward contours in the vertex part of
  $\bSigma(\cdot,\wb_j)$ explicitly only within this particular factor
  [\ie\ in the interaction propagator, associated $\coth + \tanh$
  functions, and associated Cooperons of $\bSigma(\cdot,\wb_j)$].
  Since the accociated contribution is dominated by frequencies $\wb
  \simeq 1/t$, which are small in the long-time limit, all other
  factors $\bcC (\cdot)$ and $\bSigma (\cdot, \wb_{i\neq j})$ of the
  diagram to which this $\bomega_j$-dependence has propagated may be
  Taylor-expanded in $\bomega_j$. Moreover, only the zeroth-order
  terms of this Taylor expansion need to be retained, since the others
  contain higher powers of $\bomega_j \sim 1/t$, and hence produce
  contributions with a subleading time dependence [as illustrated in
  App.~\ref{app:checkF1}, where such an expansion is carried out
  explicitly in a very similar context].

Having clarified that a truncation scheme is justified in principle in
the long-time limit, we have to implement one in practice, in such a
way that the leading terms are not affected.  In the present context,
the simplest version of such a truncation scheme would be to replace
the $\delta (\Omega_{14} + 2 \bomega)$ in
\Eq{eq:Bethe-Salpeter-new-selfenergy} by $\delta (\Omega_{14})$,
thereby rendering the entire equation for $\bSigma{}^{\cal E}_{\bmq,
  \bare} (\Omega_1, \Omega_4)$ proportional to $\delta (\Omega_{14})$.
As a result, the arguments of the ``self-energy-diagram-only''
discussion in Section~\ref{sec:standardDyson} would apply: the
Bethe-Salpter equation could then be simplified to a Dyson-type
equation, whose self-energy would be given by an expression analogous
to \Eq{eq:effectiveCooperonselfenergy}, but now including a vertex
contribution:
\begin{eqnarray}
\label{eq:effectiveCooperonselfenergy+vertex}
\bSigma{}^{\selfvertex}_{\ve , \bmq}  (\omega)
\equiv {1 \over \hbar} \int (d \bomega) (d \bbmq) \, 
\left[ \bSigma{}^{\ve , \self}_{\bmq, \bbmq, \bare} 
  + 
  \bSigma{}^{\ve , \vertex}_{\bmq, \bbmq, \bare} \right] 
  (\omega , \bomega) 
\, . \nonumber
\end{eqnarray}
However, note that it would now \emph{not} be possible to calculate
the decoherence rate $\gamma_{\ve, 0 }^{\varphi,\self} $ according to
\Eq{eq:definegammaphi} by setting $E_\bmq = i \omega $ and $\bmq = 0$
in $\bSigma{}^{\selfvertex}_{\ve , \bmq} (\omega)$, because the factor
$\bcC^0_{\bmq - \bbmq} (\omega)$ contained in $ \bSigma{}^{\ve ,
  \vertex}_{\bmq, \bbmq, \bare} (\omega , \bomega) $ would then yield
an infrared divergence for $\bbmq \to 0$. 

To avoid this problem, a version of the calculation has to be found in
which the condition $E_\bmq = i \omega$ is avoided, and the variables
$\bmq$ and $\omega$ are integrated over instead.  In
Section~\ref{sec:solutionBS} of the main text this is achieved by
transcribing the Bethe-Salpeter equation to the position-time domain,
and solving it with an exponential Ansatz, $ \tilde C (\bmr_{12}; t_1,
t_2) \simeq \tilde C^0 e^{[\tilde C^{(1)}/\tilde C^0]}$, where $
\tilde C^{(1)} (\bmr_{12}; t_1, t_2) $ is an appropriately
Fourier-transformed version of \Eq{eq:Cwbomega1} involving $\int (dq)
(dw)$ integrals, precisely as desired.  This is a factorization
approximation, in the sense that a proliferation of entangled
frequencies is avoided by approximating the $n$-th order contribution
to the Cooperon by $\tilde C^{(n)} \simeq {1 \over n!}  [ \tilde
C^{(1)} ]^n/ [\tilde C^0]^{n-1}$.

Note that in this scheme, the frequency transfer between forward and
backward lines generated by the vertex terms \emph{is treated exactly
  in the first order terms} needed for $C^{(1)}$; the factorization
approximation sets in only in second and higher orders.  Treating the
first order terms exactly is the best one can do in our
reexponentiation-of-$C^{(1)}$ scheme, since in the latter, an accurate
treatment of effects occuring only in second or higher order is beyond
the accuracy of the method.  The accumulation of energy transfers is
such an effect, but fortunately it produces corrections that are only
subleading in time, as argued above.

\section{The Bethe-Salpeter equation yields the exact Cooperon in the classical limit}
\label{app:BSAAK}

In general, solving a Bethe-Salpeter equation starting from a
self-energy calculated only to lowest order in the interaction does
not provide the exact solution of the initial problem. This remains
true even when the self-energy is treated self-consistently (i.e.
inserting the full propagators into the diagram for $\Sigma$, as we
have done for $\Sigma_{\rm full}$). Nevertheless, in the following we
shall demonstrate that \emph{for $d=1$ and classical white Nyquist
  noise}, the exact solution of the Bethe-Salpeter equation
(\ref{eq:BSpositiontau}) fully reproduces the exact results for the
Cooperon derived by AAK \cite{AAK}, implying that our Bethe-Salpeter
equation itself is exact for this type of noise. The reason for this
may be traced back to the special properties of the white Nyquist
noise interaction propagator, as will be explained below.  Thus,
non-exact results obtained from our Bethe-Salpeter equation for 
$d=1$ and white
noise are entirely due to approximations involved in constructing a
solution, such as the ``re-exponentation'' of the first order result.
(Actually, the deviations resulting from the latter approximation
produces are quantitatively rather small, as demonstrated in Paper I
[Section \ref{sec:1Dcompare}].)

We start from the Bethe-Salpeter equation,
Eq.~(\ref{eq:BSpositiontau}), using the full (not bare!)  self-energy
$\tilde \Sigma_\full$, given by \Eqs{eq:CSigmaetildewtaut4prime} and
(\ref{subeq:newselfenergies}).  The latter simplifies considerably for
classical white Nyquist noise, described by setting $\bcL^{R/A}_\qb
(\bomega) \mapsto 0$ and $-\toh i \bcL^K_\bbmq (\bomega) \mapsto
\bcL^\class_\qb \equiv T / (\nu D \qb^2 )$. The first replacement
implies that the self energy diagrams, and hence also the Cooperon
$\bcC^{\cal E}_{\bmq } ( \Omega_1 , \Omega_2) $, no longer depend on
${\cal E}$, so that this argument will be dropped henceforth.  The
second replacement results in an interaction propagator that is not
only independent of $\ve$, $\omega$ but also of $\bomega$, \ie\ white
in frequency transfers, implying that it becomes a $\delta$-function
in the time domain.  The resulting self energy has the form
\begin{eqnarray}
  \label{eq:sigmarttfullexplicit}
   \tilde \Sigma^{\ve, t^\prime_4}_\full ({\bmr_{14}} ; t_1,t_4) 
=    \WKbare (r_{14}) 
\bigl[ \delta (\tildet_{14}) - \delta (t_{14}) \bigr] \delta (t_{4'})
 , \qph
\end{eqnarray}
where $\WKbare (r_{14}) = {2 \over \hbar} \int (d \qb)
\bcL^\class_\bbmq e^{i \qb r_{14}}$.  This is ultraviolet convergent
only for $d=1$, to which we henceforth restrict our attention, but
then $\WKbare (r_{14})$ is infrared divergent. However, when
\Eq{eq:sigmarttfullexplicit} is inserted into
Eq.~(\ref{eq:BSpositiontau}), this divergence can be arranged to
cancel between the two terms of \Eq{eq:sigmarttfullexplicit}: its
second term, which stems from $\Sigma^\self$, produces [using
$\tC(r_{14}; t_1, t_1) = \delta (r_{14})$] a contribution $- \WKbare
(0) \, \tC(r_{12}; t_1, t_2)$, which can be rewritten as $- \WKbare
(0) \int d \bmr_4 {\tilde C}(r_{14}; t_1, -t_1) {\tilde C}(r_{42};
-t_1, t_2)$, so that it takes a form similar to that resulting from
the first (vertex) term of \Eq{eq:sigmarttfullexplicit}. Thus, the
Bethe-Salpeter equation (\ref{eq:BSpositiontau}) can be written as
\begin{eqnarray}
\nonumber
 &  & \bigl( - D \nabla^2_{\bmr_1} + \partial_{t_1} +  \gammaH \bigr) \, 
{\tilde C}(r_{12}; t_1, t_2)
 =  
\delta (\bmr_{12}) \, \delta (t_{12})  + 
\\
& & 
\int d \bmr_4 \WK(r_{14}) \, 
 {\tilde C}(r_{14}; t_1, -t_1) {\tilde C}(r_{42}; -t_1, t_2)   \; , 
  \label{eq:BSclassical}
\end{eqnarray}
where the kernel $\WK (r_{14}) = \WKbare (r_{14}) - \WKbare (0) $ 
is free of infrared problems:
\begin{eqnarray}
  \WK(r) = \int {d\qb \over 2 \pi}
  {4 e^2 T \over \hbar \sigma_1 }{e^{i \qb r}-1 \over \qb^2}\, .
\label{spacekernel} 
\end{eqnarray}
(Here $\sigma_1 = a^2 2 \nu e^2 D$ is the inverse resistance per
length of a quasi 1-dimensional wire of cross section $a^2$.)
\begin{figure}
\begin{center}\includegraphics[%
  width=0.90\columnwidth]{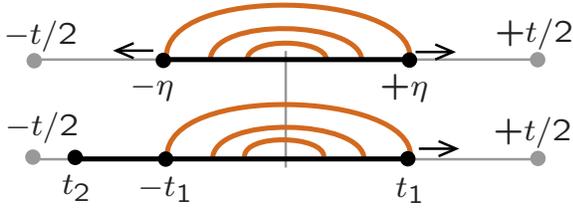}
\end{center}
\caption{\label{cap:AAKCooperon}
Schematic illustration of the time-evolution in the differential
equation for the Cooperon: AAK's scheme (top) evolves both endpoints
symmetrically, while the full Bethe-Salpeter equation (bottom) contains two
Cooperons, describing propagation in the intervals $(t_2,-t_1)$ and
$(-t_1,t_1)$. Curved lines represent interaction propagators.  
}
\end{figure}

Eq. (\ref{eq:BSclassical}) is solved by the following path integral:
\begin{eqnarray}
&&{\tilde C}(r_{12};t_1,t_2)=\theta(t_1-t_2) \nonumber \\
&&\times \int {\mathcal D}r \exp \left[ 
-{\gamma_H} (t_1-t_2)- \int_{t_2}^{t_1} {{\dot r}^2(t') \over 4D} dt'
 \right.
\nonumber\\
&& \left. + \int_0^{t^*_1} dt' \WK\bigl(r(t')-r(-t')\bigr) \right] 
\label{BSpathintegral} \, ,
\end{eqnarray}
where $t_1^*=t_1$ for $-t_2>t_1>0$, $t_1^*=-t_2$ for $t_1>-t_2>0$ and
$t_1^*=0$ otherwise. Indeed, upon differentiating with respect to
$t_1$, we find that the contribution from the
$\WK$-term in the path-integral is nonzero only
for $0 < t_1 < -t_2 $, as is the case for the
$\WK$-term in Eq.~(\ref{eq:BSclassical}), and in fact
precisely equals the latter:
\begin{eqnarray}
  && \int {\mathcal D}r \WK(r(t_1)-r(-t_1)) \exp[\ldots] \nonumber \\
  &=& \int dr_4 \WK(r_{14}) {\tilde C}(r_{14}; t_1, -t_1) {\tilde C}(r_{42}; -t_1, t_2) \, .
\label{pathcontribution}
\end{eqnarray} 
The validity of this equality would be obvious if we were treating
unmodified Cooperons. The fact that it remains true even in the
presence of noise is due to the special nature of this noise: In
principle, we need a ``three-point Cooperon'' involving times $t_1,
-t_1$, and $t_2$, and it factorizes in the manner shown in
\Eq{pathcontribution} only because the classical Nyquist noise
correlator is a $\delta$-function in time: indeed, as can be seen from
Eq.  (\ref{eq:BSclassical}) itself, the noise correlator always only
connects time-points $t'$ and $-t'$, which means there is no
correlator connecting two points in the disjoint intervals
$(t_2,-t_1)$ and $(-t_1,t_1)$ involved here (see
Fig.~\ref{cap:AAKCooperon}, bottom).




By rewriting the path-integral (\ref{BSpathintegral}) for the special
case of equal times $t_2=-\eta$, $t_1=\eta$, it can be shown to be
identical to the exact Cooperon path-integral expression considered by
Altshuler, Aronov and Khmelnitskii in their seminal work (Eq.~(20) in
Ref. \onlinecite{AAK}, corrected\cite{factorof2} for factors of 2),
where ${\tilde C}(0;\eta,-\eta)=\tau_{\rm el} C^{\rm
  AAK}_{\eta,-\eta}(r,r)$,
\begin{eqnarray}
\int_0^{\eta} dt' 
\WK (r(t')-r(-t')) = \hspace{3cm} \qqph \\
- {2 e^2 T\over \hbar \sigma_1} 
\int_{-\eta}^{\eta} dt' \int_{-\infty}^{+\infty} {dk \over 2
  \pi k^2} \Bigl\{1-\cos\bigl(k[r(t')-r(-t')]\bigr) \Bigr\}  
\nonumber 
\end{eqnarray}
In comparing the expressions, note that our $\gamma_H$ plays the role
of their $1/\tau_{\varphi}$ as a given extrinsic decoherence rate,
and our $\tauphi$ the role of their $\tau_N$.

Finally, we comment on the connection between our Bethe-Salpeter
equation (\ref{eq:BSclassical}), which is quadratic in the Cooperon,
and AAK's differential equation for the Cooperon (Eq.~(23) in
Ref.~\onlinecite{AAK}, corrected\cite{factorof2} for factors of 2),
which is linear:
\begin{eqnarray}
\left( {\partial \over \partial \eta} - D {\partial^2 \over \partial
    \rho^2} + {2 \gamma_H} + {2 \sqrt{2} e^2 T\over \hbar \sigma_1}
  |\rho|\right) C^{\rm AAK}_{\eta,-\eta}(\rho,\rho')= & & \nonumber \\
{\delta(\eta)\delta(\rho-\rho') \over \sqrt{2} \tau_{\rm el}} \; \; && 
\label{eq:AAKC}
\end{eqnarray}
Here $\rho(t)=[r(t)-r(-t)]/\sqrt{2}$ is the difference coordinate
introduced by AAK. Both equations yield the same result, since they
are solved by the same exact path integral. The origin of the
difference between the two equations is that \Eqs{eq:BSclassical} and
(\ref{eq:AAKC}) is that AAK's \Eq{eq:AAKC} describes the symmetric
evolution of both the first and second time arguments of their
Cooperon, namely $\eta$ and $- \eta$ (as illustrated in
Fig.~\ref{cap:AAKCooperon}, top), wheras in our \Eq{eq:BSclassical},
only the first time argument of the Cooperon, $t_1$, is time-evolved.
For $\eta=t_1=-t_2$, an integral such as the one in the second line of
our equation (\ref{eq:BSclassical}) becomes linear in $\tilde C$,
since the second Cooperon reduces to a $\delta (r_{24}$ function.
However, such a simplification is possible only if the interaction
propagator is a $\delta$-function in time, and therefore cannot be
employed to simplify evaluation of the full Bethe-Salpeter equation in
general, with interaction propagators more long-ranged in time.


\section{Some subleading corrections 
to $\tilde F^\ve$ and  $\tilde C^{(1) \ve}$}

In this appendix we calculate some subleading corrections to the decay
function $\tilde F^\ve$ and the first order Cooperon $\tilde C^{(1)
  \ve}$, which were mentioned but not discussed in detail in the main
text. 

\subsection{Calculation of $\tilde F_1^\ve$}
\label{app:checkF1}

In  \Eq{eq:Fonefirstresults} in Section~\ref{sec:evaluatingFe},
we expanded the propagator $\bar {\cal L}^{\deco}_{\ve \omega, \bbmq}
(\bomega)$ in powers of $\omega$, and subsequently evaluated only the
corresponding lowest order contribution $\tilde F^\ve_0$ to the decay
function, arguing that small $\omega$ dominate in the long-time limit
so that the higher terms are negligible.  Let us now check this
explicitly by calculating the first correction, $\tilde 
F^{(1)}_d (t) \equiv \langle
\tilde F_{(1)}^\ve (t) \rangle_\ve$, starting from \Eqs{eq:Fonenresults},
and using methods and notations analagous to those of Paper I,
Sections~\ref{eq:generalF} and~\ref{sec:e-averaged}.
It will be found to be subleading, so we shall only calculate its
order of magnitude, without caring about numerical prefactors.
We begin by noting that 
\Eq{eq:define:Ldeco:n} yields
$  \langle \bcL^{\deco}_{\ve (n), \bbmq} (\bomega) \rangle_\ve = 
- {\cal W}_{\app}^{(1)} (\wb) / \nu$ , where 
\begin{eqnarray}
  \label{eq:L(1)}
{\cal W}_{\app}^{(1)} (\wb) = {\textstyle {1 \over 8 T}}
\langle \textrm{sech}^2 \bigl[ (\ve - \wb)/2T \bigr] \rangle_\ve \; .
\end{eqnarray}
Writing its Fourier transform as 
\begin{eqnarray}
  \label{eq:w1(z)}
  w^{(1)}_\app (t_{34} T) \equiv 
  \int (d \wb) e^{-i \wb t_{34}} {\cal W}_{\app}^{(1)} (\wb) \; , 
\end{eqnarray}
we note that the function $  w^{(1)}_\app (z)$ is dimensionless, 
peaked around zero, with heigth, width and weight all of order 1.
Inserting this into \Eq{eq:Fonenresults} for $n=1$, writing
the latter in a form similar to \Eq{eq:exponentBetheSalpeter},
and rewriting the time integrals in terms
of the dimensionless sum and difference variables
$\tilde x = \tilde t_{34}/t$ and 
$z = t_{34}T = xTt$, with $x = t_{34}/t$, 
we obtain
\begin{eqnarray}
  \label{eq:F(1)explicit1}
  \tilde F^{(1)}_d (t) = {D_t \over T}
  \! \int_0^{Tt}  \!\! d z \, w_\app^{(1)} (z) 
  \, {\cal P}^{(1)}_d (z/Tt)   \; , 
\end{eqnarray}
where we defined the operator $D_t = (- \partial_t + d/(2t) +
\gammaH)$, the function ${\cal P}^{(1)}_d (x) = \int_{x_0}^{1 - x} \!\!  d
\tilde x \, \delta \tP_d^{(1)} ( \tau,\ttau) $, with
\begin{eqnarray}
\nonumber
  \delta \tP_d^{(1)} ( \tau,\ttau)
& = & {2 t \over \hbar \nu }  \int (d \bbmq) \left[
\bP^\crw_{(0, t)}(\qb , |t_{34}|)  -
\bP^\crw_{(0, t)}(\qb , |\tildet_{34}|)  
 \right] 
\\
  \label{eq:definedeltaP1d}
& = & 
{2^{2-d} \over \pi^{d/2} g_d (L_t)} \left[
\tau^{-d/2} - \ttau^{-d/2}  \right] \; , 
\end{eqnarray}
and the shorthand notations $\tau = (1-x)x$, $\ttau = (1-\tilde
x)\tilde x$. Moreover, we introduced an ultraviolet cutoff $x_0 =
t_0/t$ in \Eq{eq:F(1)explicit1}, which will be needed for $d=2,3$, and
$t_0$ can be taken as the elastic scattering time $\tauel$.  The fact
that a cutoff is needed is of no great concern, since the leading
long-time behavior will turn out to be subdominant anyway.

In the limit $Tt \gg 1$, we need only the asymptotic small-$x$
behavior of ${\cal P}_d (x)$ in \Eq{eq:F(1)explicit1}, which 
is given by 
\begin{eqnarray}
  \label{eq:calPdexplicit}
  {\cal P}^{(1)}_1 (x) & = & 
  {2 \over \pi^{1/2} g_1 (L_t)} 
  \Bigl[ x^{-1/2} + \dots \Bigr] \; , 
  \\  
  {\cal P}_2^{(1)} (x) & = & 
  {1 \over \pi g_2 (L_t) } 
  \Bigl[ x^{-1} + \ln (x x_0) + \dots \Bigr] \; , 
  \\
  {\cal P}_3^{(1)} (x) & = & 
  {1 \over 2 \pi^{3/2} g_3 (L_t) }  
  \Bigl[ x^{-3/2} - 2 x_0^{-1/2} + \dots \Bigr] \; . \qqph 
\end{eqnarray}
Inserting this into \Eq{eq:F(1)explicit1}, we find that $\tilde F^{(1)}_d
(t)$ is smaller than the leading term $\tilde F^\app_d (t) \equiv = \langle
\tilde F^\ve_{(0)} (t) \rangle_\ve$, given by \EqsI{subeq:F(t)g(L_t)} of
paper I, by powers of the parameters $\tgammat = 1/Tt$ and $\tgammaH =
1/T \tauH$, which for present purposes are both $\ll 1$:
\begin{eqnarray}
  \label{eq:Fpp(1)/overFpp}
  \tilde F_1^{(1)} (t) & = & \tilde F_1^\app (t) \, 
  {\cal O} \Bigl[\tgammaH \tgammat^{1/2},  \tgammat^{3/2} \Bigr]  , 
  \\
  \tilde F_2^{(1)} (t) & = & {\tilde F_2^\app (t) 
    \over \ln (Tt)} \, 
  {\cal O} \Bigl[\tgammaH , (\tgammaH \tgammat , \tgammat^2 ) 
  \ln (T t_0) \Bigr]  , \qqph  
  \\
  \tilde F_3^{(1)} (t) & = & \tilde F_3^\app (t) \, 
  {\cal O} \Bigl[\tgammaH, \tgammat, (\tgammaH \tgammat, \tgammat^2)/
  (T t_0)^{1/2}  \Bigr]    .  \qqph 
\end{eqnarray}
In particular, recalling the leading time dependence of
$\tilde F^\app_d (t)$ [namely $t^{3/2}$, $t \ln (Tt)$ or $t$ for $d =1,2,3$], 
we see that the leading terms
of $\tilde F^{(1)}_d (t)$ all either vanish in the limit of
no magnetic field ($\tgammaH = 0$), or are constant
or decreasing functions of time. Hence we conclude that
$\tilde F^{(1)}_d (t)$ indeed can be neglected for the purposes
of determining the decoherence time. \\

\subsection{Long-time behaviour of $\tilde{C}^{(1) \ve}_{\rm self,Z}$}
\label{sec:SigmaZsubleading}

Next, we consider in more detail the correction $\tilde{C}^{(1)
  \ve}_{\self, Z}(t) \equiv \tilde{C}^{(1) \ve}_{\self, Z} (0; \toh t,
- \toh t)$ to the Cooperon arising from $\bSigma^{\self ,Z}$; it is
given by an equation similar to the first term of
(\ref{eq:C1tauexplicit}), but using $\bar \Sigma^{\self, Z}$
[\Eq{eq:SigmabareZ}] as self-energy.  This contribution was
purposefully \emph{omitted} in our calculation of $\tilde{C}^{(1)
  \ve}_\deco (t) \equiv \tilde{C}^{(1) \ve}_\deco(0; \toh t, - \toh
t)$ from (\ref{eq:C1tauexplicit}).  To justify this omission, we shall
now show that the long-time behaviour of $\tilde{C}^{(1) \ve}_{\self
  ,Z}(t)$ is subdominant as compared to the leading behaviour from
$\bigl \langle \tilde C^{(1) \ve}_\deco (t) \bigr \rangle_\ve$.

It turns out that for this
calculation, we have to introduce both an IR cutoff in $\bar{q}$, 
and an UV cutoff in $\bomega$
[though the latter would not be necessary if, in contrast to the
calculation below, one would use the general expression
(\ref{eq:generalinteraction}) for the interaction propagator instead
of the unitary limit (\ref{eq:recallLCD})].  The occurence of these
divergencies is not a surprise, since the self-energy diagrams used in
our Bethe-Salpeter equation constitute only a subset of the diagrams
that make up the cross-terms of interaction- and weak-localization
corrections to the conductivity, namely that subset of diagrams
capable of being iterated in a diagrammatic equation for the Cooperon.
However, it has been shown in Ref.~\onlinecite{AAG} that when
\emph{all} contributions to the conductivity to first order in the
interaction (and second order in $1/g$) are calculated, numerous
additional terms arise which turn out to cancel the abovementioned IR
and UV divergencies, but which we have not considered here.

For present purposes, it is sufficient to simply cut off these
divergence by hand, since we shall find that the leading
  long-time behavior of this term is subdominant anyway.  To identify
this long-time behavior, we shall isolate the strongest singularity in
the frequency domain of the Fourier transform $\tilde {\cal
  C}^{(1) \ve}_{\self ,Z} 
(\omega) \equiv \int d t \, e^{i \omega t}
\tilde {C}^{(1) \ve}_{\self ,Z} (t)$, 
which has the following form:
\begin{widetext}
\begin{subequations}
\begin{eqnarray}
  \tilde {\cal C}^{(1) \ve}_{\self ,Z}(\omega) 
  & = & \oneoverhbar \int(d\bbmq)(d\bomega)\,
  \frac{i}{2}  \bLRbbqw  \left [\tanh(\frac{\ve-\bomega}{2T})-
  \tanh(\frac{\ve+\bomega-\omega}{2T}) \right]\, 
  I[\bar{q},\bar{\omega},\omega] \; , 
\label{appselfZ_CselfZdef}
\\
I[\bar{q},\bar{\omega},\omega] & = & \int(dq) \, \bar{{\cal
    C}}_{q}^{0}(\omega)^{2}
\left[\left[\bar{{\cal
        D}}_{\bbmq}^{0}(\bomega)\right]^{2}\left(\left[\bar{{\cal
          C}}_{q}^{0}(\omega)\right]^{-1}+\left[\bar{{\cal
          D}}_{\bbmq}^{0}(\bomega)\right]^{-1}\right)-\bar{{\cal
      C}}_{q-\bbmq}^{0}(\omega+\bomega) \right] \; . 
\label{appselfZ_I}\end{eqnarray}
\end{subequations}
We shall analyse these expressions explicitly only for $d =1$, which
is most prone to infrared divergencies for $\omega \to 0$, which would
correspond to a long-time behavior that grows with time $t$; if
$\tilde {C}^{(1) \ve}_{\self ,Z} (t)$ is found to be subleading for
$d=1$, the same will be true for $d=2,3$.  Using \Eq{eq:recallLCD} for
$\bLRbbqw$, the contour integral over $q$ yields $I=(I_{1}-I_{2})/a^2$
{[}referring to the first and second terms in square brackets{]}, with
\begin{subequations}
\begin{eqnarray}
I_{1} & = & \frac{1}{4\sqrt{D}}
\frac{2\gamma-i(2\omega+\bar{\omega})+
D\bar{q}^{2}}{(\gamma-i\omega)^{3/2}(-i\bar{\omega}
+D\bar{q}^{2})^{2}} \; , \qqph 
\label{appselfZ_I1} 
\\
I_{2} & = & \frac{i}{\sqrt{D} (2z_{2})[(\sqrt{D}\bar{q}+
z_{2})^{2}-z_{1}^{2}]^{2}}- 
\frac{i}{\sqrt{D} (2z_{1})^{2}[(z_{1}-
\sqrt{D}\bar{q})^{2}-z_{2}^{2}]}
\left[\frac{1}{z_{1}}+\frac{2(z_{1}-
\sqrt{D}\bar{q})}{(z_{1}-\sqrt{D}\bar{q})^{2}-z_{2}^{2}}\right]
\,, \qqph
\label{appselfZ_I2}
\end{eqnarray}
\end{subequations}
\end{widetext}
where $z_{1}=i\sqrt{\gamma_{H}-i\omega}$ and 
$z_{2}=i\sqrt{\gamma_{H}-i(\omega+\bar{\omega})}$.

For $\gamma_{H}\rightarrow0^{+}$, the part of $I$ which 
will yield the most singular contribution in the 
frequency domain (after
integration over $\bar{\omega}$ and $\bar{q}$)
is a $1/\sqrt{\omega}$ singularity at $\omega=0$
($I_{1,2}$ both yield a $\omega^{-3/2}$ contribution, but those
cancel in $I_{1}-I_{2}$):
\begin{equation}
  I[\bar{q},\bar{\omega},\omega]\approx-
  \frac{\bar{\omega}/\sqrt{D}}{a^2(\bar{\omega}+iD\bar{q}^{2})^{3}}
  \frac{1}{\sqrt{0^{+}-i\omega}}\,.\label{appselfZ_I_leading}\end{equation}
The occurence of a $\omega^{-1/2}$ singularity can be understood as
follows: for $\omega = 0$, the integrand in $I[\qb, \wb, 0]$ diverges
as $1/(D q^2)$ (for $d=1$), but this divergence is cut off by $\omega
\neq 0 $, so that the integral goes as  $I[\qb, \wb, \omega] \sim
\omega^{-1/2}$. (By a similar argument, it follows that for $ d = 2,3$,
the leading $ \omega$ dependence will be less singular,
namely $\ln (\omega)$ or a constant, respectively).

The subsequent integrals of this term over $\bar{q}$ and 
$\bar{\omega}$ need an IR- and UV-cutoff $\bar{q}_{{\rm min}}$
and $\bar{\omega}_{{\rm max}}$, respectively. Keeping only
the contribution that dominates for $\bar{q}_{{\rm min}}\rightarrow0$ and 
$\bar{\omega}_{{\rm max}}\rightarrow\infty$,
we find:
\begin{equation}
  \bar{C}^{(1) \ve}_{\self ,Z} (\omega)
  \sim\frac{2}{\hbar a^4 \nu D\sqrt{0^{+}-i\omega}}
  \frac{\ln\left(\frac{{\rm max}(\ve,T)}{\bar{\omega}_{{\rm
            max}}}\right)}
  {\sqrt{D}\bar{q}_{{\rm min}}} \; . 
\label{appselfZ_CselfZleading}
\end{equation}
 The corresponding temporal behaviour is
\begin{eqnarray}
\label{eq:C1fromSigmaZ}
\tilde{C}^{(1) \ve}_{\self
  ,Z}(t) \sim {t^{-1/2} \over \hbar a^4 \nu D}
  \frac{\ln\left(\frac{{\rm max}(\ve,T)}{\bar{\omega}_{{\rm
            max}}}\right)}
  {\sqrt{D}\bar{q}_{{\rm min}}} 
 \; , 
\end{eqnarray} which is subleading at long times compared to
$\tilde C^{(1) \ve}_\deco (t) \sim t^{1}$ 
[from \Eqs{eq:finalresultforEFfull} and
(\ref{subeq:linearizedFequation})].
Even upon inserting
$\bar{q}_{{\rm min}}\sim (Dt)^{-1/2}$ for the IR cutoff, 
so that $\tilde{C}^{(1) \ve}_{\self ,Z}(t) \sim {1/(\hbar a^4 \nu D)}$,
we note that  as long as $g_1 (L_\varphi) \gg 1$, 
this contribution is much smaller than the
leading contribution to $\tC^{(1),\ve} (t)$ at scales
$t \simeq \tauphi$,  namely
$\tilde C^{(1) \ve}_\deco 
(\tauphi) \simeq a^{-2} (D\tauphi)^{-1/2} =g_1 (L_\varphi) / 
(\hbar a^4 \nu D) $.


\begin{thebibliography}{10}
\bibitem{gammaH/T}
To obtain \Eq{eq:cancelspontabs} from 
\Eq{eq:define:Ldeco} for $\bcL^{\deco}_{{\cal E} \omega, \bbmq}
(\bomega)$,  we should replace the $\omega/T$ in the
second $\tanh$ function by $E_q/T$, which in the limit $q\to 0$
reduces to $\gammaH/T$. For
not too large magnetic fields, this is $\ll 1$ and 
can be neglected.
\bibitem{factorof2} Ref.~\onlinecite{AAK} of AAK is known to contain
  various incorrect factors of two.  They can be eliminated by
  replacing $\eta/2 \to \eta$ in AAK's Eq.~(6), and rederiving all
  subsequent equations. For example, one needs $\toh \omega (t_1 \pm
  t_2) \to \omega (t_1 \pm t_2)$ in Eq.~(9) and $T \to 2T$ in
  Eqs.~(19) and subsequent equations, \eg\ (20), (22), (23), (26) and
  (28), so that their $d=1$ decoherence time ($\tau_N$) and
  magnetoconductivity ($\Delta \sigma$) take the forms given in this
  paper (or by Eqs.~(2.38b) and (2.42c) of Ref.~\onlinecite{AAG},
  respectively).
\end{thebibliography}
\end{document}